\documentclass[showpacs,reprint,twocolumn,preprintnumbers,amsmath,amssymb,aps,superscriptaddress]{revtex4-1}

\usepackage{graphicx}
\usepackage{dcolumn}
\usepackage{bm}
\usepackage{subfigure}
\usepackage{color}
\usepackage{times}

\renewcommand{\eqref}[1]{(\ref{#1})}

\begin{document}
\title{Quantum Enhanced Measurement of Rotations with a Spin-1 Bose-Einstein Condensate in a Ring Trap}
\author{Samuel~P.~Nolan}
\email{uqsnolan@uq.edu.au}
\affiliation{School of Mathematics and Physics, The University of Queensland, Brisbane, Queensland, Australia}
\author{Jacopo~Sabbatini}
\affiliation{School of Mathematics and Physics, The University of Queensland, Brisbane, Queensland, Australia}
\affiliation{ARC Centre of Excellence for Engineered Quantum Systems, The University of Queensland, Brisbane, Queensland, Australia}
\author{Michael~W.~J.~Bromley}
\affiliation{School of Mathematics and Physics, The University of Queensland, Brisbane, Queensland, Australia}
 \author{Matthew ~J.~Davis}
\affiliation{School of Mathematics and Physics, The University of Queensland, Brisbane, Queensland, Australia}
\affiliation{JILA, University of Colorado, 440 UCB, Boulder, Colorado 80309, USA}
\author{Simon~A.~Haine}
\affiliation{School of Mathematics and Physics, The University of Queensland, Brisbane, Queensland, Australia}

\date{\today}

\begin{abstract}

We present a model of a spin-squeezed rotation sensor utilising the Sagnac effect in a spin-1 Bose-Einstein condensate in a ring trap. The two input states for the interferometer are seeded using  Raman pulses with Laguerre-Gauss beams and are amplified by the bosonic enhancement of spin-exchange collisions, resulting in spin-squeezing and potential quantum enhancement in the interferometry.  The ring geometry has an advantage over separated beam path atomic rotation sensors due to the uniform condensate density. We model the interferometer both analytically and numerically for realistic experimental parameters and find that significant quantum enhancement is possible, but this enhancement is partially degraded when working in a regime with strong atomic interactions.

\end{abstract}


\maketitle

\section{Introduction}\label{sec1} 

Atom interferometers are relatively new measurement devices that harness the wave nature of atoms at low temperatures to measure quantities such as magnetic fields \cite{Vengalattore2007, Muessel2014} and physical constants \cite{Fixler2007,Bouchendira2011} with ever increasing precision. In particular, matter-wave interferometry is particularly sensitive to  inertial measurements such as gravitational fields \cite{peters1999,peters2001,McGuirk2002, Altin2013} and rotations \cite{Gustavson1997, Gustavson2000, Durfee2006, burke09a, dickerson13a}. Precision rotation sensing is of practical interest with applications in navigation technology and geophysics, and it may also play an important role in the detection of gravitational waves \cite{Lantz2009}.

At nanokelvin temperatures, atomic Bose-Einstein condensates (BEC) provide a near monochromatic source of matter waves, which can potentially lead to improved visibility and decreased uncertainities in interferometric experiments as compared to laser-cooled thermal atoms \cite{Altin2011a, Debs2011, Szigeti2012, Hardman2014}. However, a major practical limitation is a reduced particle number available for the interferometer as compared to laser-cooled gases. \cite{Robins2005,Robins2013}. 

The minimum phase uncertainty that can be achieved with an atom interferometer using uncorrelated sources is the standard quantum limit (SQL), $\Delta \phi = 1/\sqrt{N_{\rm t}}$ where $N_{\rm t}$ is the total number of atoms used in the experiment \cite{Wineland1994, Dowling1998}. Because of this limitation, it is desirable to devise schemes that are able to boost phase sensitivity without requiring more atoms.  The performance of atom interferometers can potentially be enhanced beyond the SQL using the method of spin-squeezing to generate correlated atomic sources. The maximum sensitivity of such sources is known as the Heisenberg limit (HL), $\Delta \phi = 1/N_{\rm t}$ \cite{Holland1993}.  

In the past two decades there have been many proposals to generate spin-squeezed states in atomic systems.  These include one-axis and two-axis twisting \cite{Kitagawa1993, Sorensen2002a, Johnsson2007a, Li2008, Liu2011, Haine2009, He2012, Haine2014, Huang2015}, molecular dissociation \cite{Kheruntsyan2005}, four-wave mixing \cite{Haine2011, Bouchoule2012} and spin-exchange collisions \cite{Duan2002, Cirac2000, Gabbrielli2015}. Of these possibilities, one-axis twisting \cite{Riedel2010, Gross2010, Leroux2010}, four-wave mixing \cite{Dall2009, Bucker2011} and spin-exchange collisions \cite{Lucke2011, Gross2011, Bookjans2011, Hamley2012} have all been demonstrated experimentally. However, to date a spin-squeezed, separated beam path interferometer, required to measure inertial effects, has not been realised.

A significant obstacle to performing spin-squeezed separated beam path interferometry with a BEC is mode-matching: in order to observe the high-contrast interference fringes required for sub-SQL interferometry, the two wave-packets to be interfered must have similar spatial density and phase profiles. Typically in a separated beam path interferometer, two atomic matter-wave packets begin as identical copies, which then traverse separate spatial trajectories before being recombined. Atomic interactions perturb the phase-profile of each wave-packet as they separate and evolve independently. Upon recombination the wave-packets will no longer overlap perfectly, which leads to reduced fringe visibility and acts essentially as signal loss, to which quantum-enhanced interferometry is highly sensitive. Additionally, phase diffusion due to the nonlinear nature of the atomic interactions is significantly increased while the clouds are not overlapped \cite{Riedel2010}, which limits the maximum
 interrogation time of the device. 

Some of these difficulties can be addressed by utilising a BEC in a toroidal trap.  This geometry results in a BEC with uniform density about the ring, which eliminates any perturbations to the phase-profile caused by wave-packet separation, and minimizes the effect of phase-diffusion caused by path separation. For this reason there have been several proposed methods for a quantum-enhanced rotation sensor constructed from a BEC in a ring trap \cite{Cooper2010, Jing2011, Jing2013, Rico-Gutierrez2013, Rico-Gutierrez2015, Opatrny2015, Das2015}.  Another proposal exploits Fermi statistics to generate correlations \cite{Cooper2012}. 

Although a spin-squeezed gyroscope has yet to be demonstrated, high-precision (but classical) gyroscopes that utilise the Sagnac effect have been realised.  These are separated beam path interferometer whereby a rotation produces a phase shift between the separated wave-packets~\cite{Gustavson1997, Gustavson2000, Durfee2006, burke09a, dickerson13a}. More recently it has also been demonstrated that a single component BEC in a ring trap can also measure rotations by exciting counter-propagating acoustic waves \cite{Marti2015}. In this paper, we investigate a rotation sensor based on a BEC uniformly filling a ring trap, and investigate how spin-exchange collisions can be used to enhance the sensitivity to better than the standard quantum limit. 

The structure of the paper is as follows. In Sec.~\ref{sec2} we outline an interferometry protocol similar to that of \cite{Halkyard2010} but which couples different spin states with Raman transitions.   We also define the relevant pseudo-spin representation and spin-squeezing parameter for the system. Sec.~\ref{sec3} provides a full description of the interferometric scheme, including the Hamiltonian and the preparation of a spin-squeezed input state.
In Sec.~\ref{sec4} the spin-squeezing of the input state is estimated analytically before a more complete numerical treatment in Sec.~\ref{sec6}. The input state is found to have sensitivity significantly below the SQL in both situations. The full interferometer sequence is simulated in Sec.~\ref{sec7} which reveals a fundamental limitation: that the squeezing parameter oscillates during the interrogation time as a result of unwanted population in other angular momentum modes due to spontaneous collisions. 

\section{Interferometric scheme}\label{sec2}

The scheme we will describe in detail below is a type of Mach-Zehnder interferometer. The key part of the interferometer is the initial equal mixing of two separate modes using an effective beam splitter, which are then allowed to freely evolve under a rotation for a certain interrogation time, before being recombined with another 50-50 beam splitting operation. The toroidal trapping geometry makes it natural to use Laguerre-Gauss (LG) beams to implement Raman transitions. We are motivated by recent work showing that  orbital angular momentum carried by the wavefront of a Laguerre-Gauss (LG) optical beam can be transferred to the centre of mass angular momentum mode of a BEC, theoretically \cite{marzlin97a, kapale05a, Sun2015, Sun2015a} and experimentally \cite{Andersen2006, wright08a, wright09a}. Our scheme utilises this idea by coupling the centre of mass angular momentum modes of a spinor BEC in a ring trap geometry, similar to Ref.~\cite{Halkyard2010}. In this section we give a broad outline of a type of interferometer that uses these Raman pulses for the beam splitting, and define the appropriate observables to measure the corresponding phase difference of the two paths.

\subsection{Heisenberg picture description of a Raman interferometer} \label{sec2a}

\begin{figure}[h!]
\centering
\begin{subfigure}
\centering
\includegraphics[width=0.8\columnwidth]{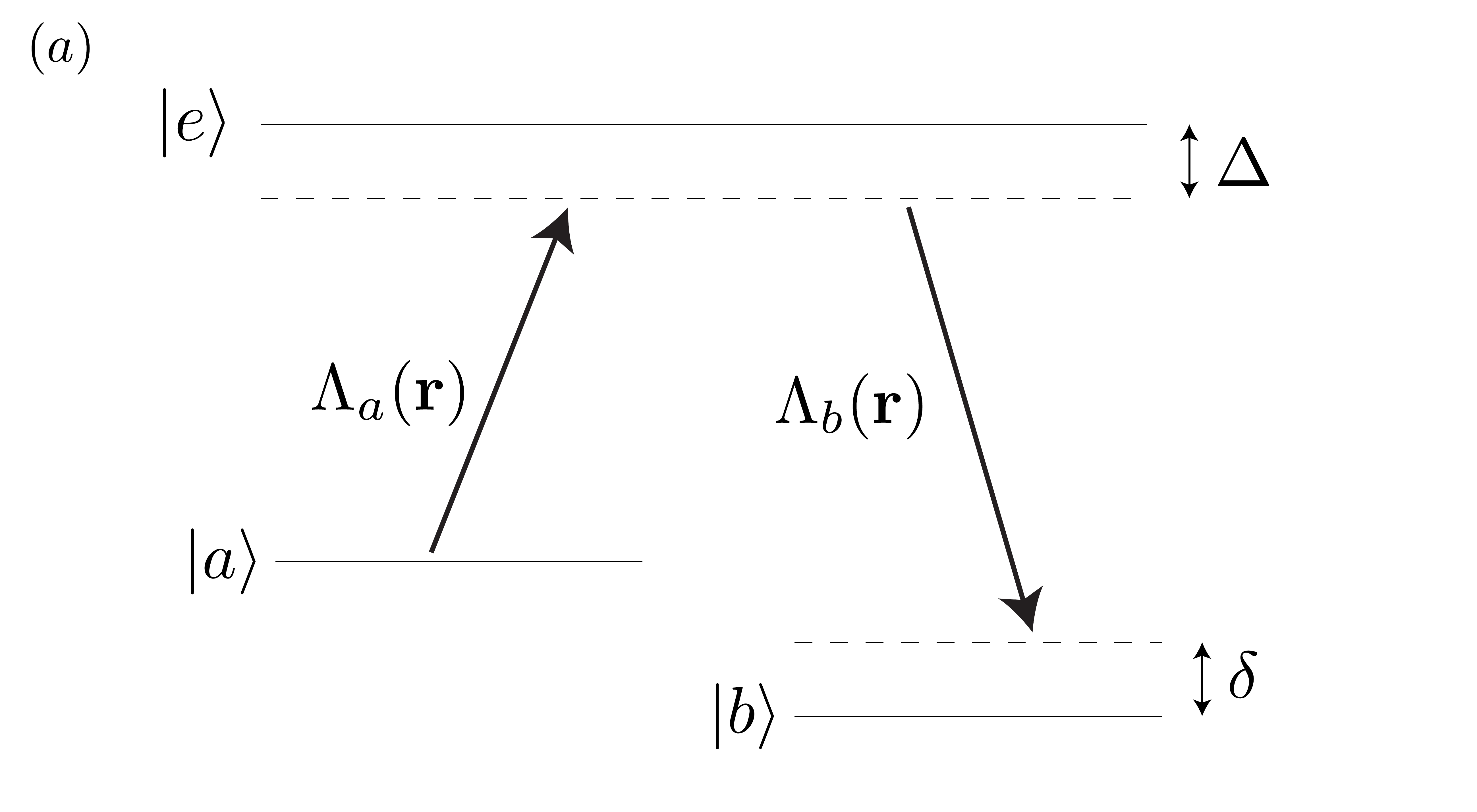}
\end{subfigure}
\hspace{2mm}
\begin{subfigure}
\centering
\includegraphics[width=0.8\columnwidth]{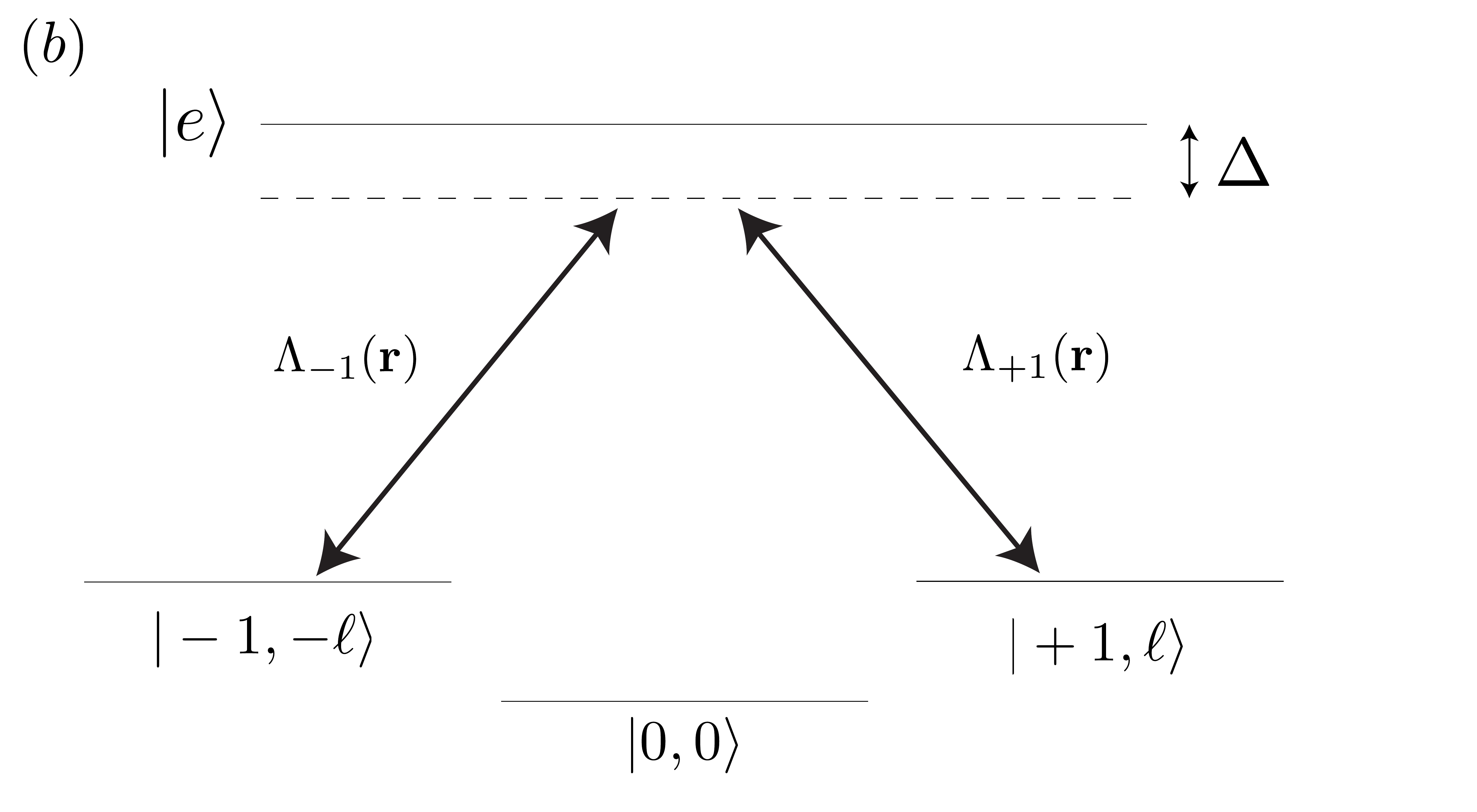}
\end{subfigure}
\qquad
\caption{ (a) An energy level diagram illustrating a general Raman transition between two spin states $|a\rangle$ and $|b\rangle$. The single photon detuning $\Delta$ is required to minimise the excited state population and the two photon detuning $\delta$ is included as a resonance condition. (b) Energy level diagram in the presence of the quadratic Zeeman shift for a $\pi/2$ pulse (``B" and ``C" in Fig.~\ref{fig:scheme}), which acts as an atomic beam splitter between the $|+1,+\ell\rangle$ and $|-1,-\ell\rangle$ modes, where we have written the atomic states as $|m_{\rm F},\ell\rangle$ where $m_{\rm F}$ is the electronic Zeeman sublevel and $\ell \hbar$ is the atomic centre of mass angular momentum mode. }
\label{fig:energylevel}
\end{figure}

A Raman transition is a well established technique in atom optics that is used to drive transitions between different electronic states of an atom while also transferring kinetic energy to the atoms \cite{Kasevich1991}, as illustrated Fig.~\ref{fig:energylevel}(a). Treating the optical beams semiclassically, making the rotating-wave approximation \cite{walls2008quantum} and adiabatically eliminating the excited state \cite{Haine2005}, the Hamiltonian which describes a two-photon Raman transition between the $m_{\rm F}=a$,$b$ Zeeman states is \cite{Szigeti2014}
\begin{multline}\label{eq:ramanHam}
\hat{\mathcal{H}}_{\rm R} = \hbar \int d\textbf{r} \left( \hat{\psi}_{a}(\textbf{r}) \hat{\psi}^\dagger_{b}(\textbf{r}) \frac{\Lambda_{a}(\textbf{r}) \Lambda^*_{b}(\textbf{r})}{2 \Delta} + \mathrm{H.c} \right) \\
+ \hbar \delta \int d \textbf{r} \ \hat{\psi}_{a}^\dagger(\textbf{r}) \hat{\psi}_{a}(\textbf{r}) ,
\end{multline}
where $\hat{\psi}_a(\textbf{r})$ is the bosonic field operator annihilating the $a$th spin state at position $\textbf{r}$, $\Delta$ is the single photon detuning frequency, H.c denotes Hermitian conjugate, $\delta$ is the two-photon detuning, and $\Lambda_{a,b}(\textbf{r})$ are the complex fields representing the LG beams. 

As well as linear momentum, LG photons carry orbital angular momentum $\hbar \ell$ where $\ell$ is an integer winding number. In cylindrical coordinates $(r, \theta, z)$  the LG beams are
\begin{equation} \label{eq:rabi}
\Lambda_{a}(\textbf{r})=\Lambda_0 e^{i { k}_{a}z }e^{i a \ell \theta},
\end{equation}
where $\Lambda_0$ is the single photon Rabi frequency between ground and excited states. We have assumed that the width of the ring trap is sufficiently small that the intensity of the LG beams is constant in this region. To couple between centre of mass motional states with orbital angular momentum $\pm \hbar \ell$ we chose the LG beams to co-propagate ($k_{+1}=k_{-1}$) with equal and opposite winding number, and assume the atoms are confined to the plane $z=0$.

\begin{figure}
\includegraphics[width=0.8\columnwidth]{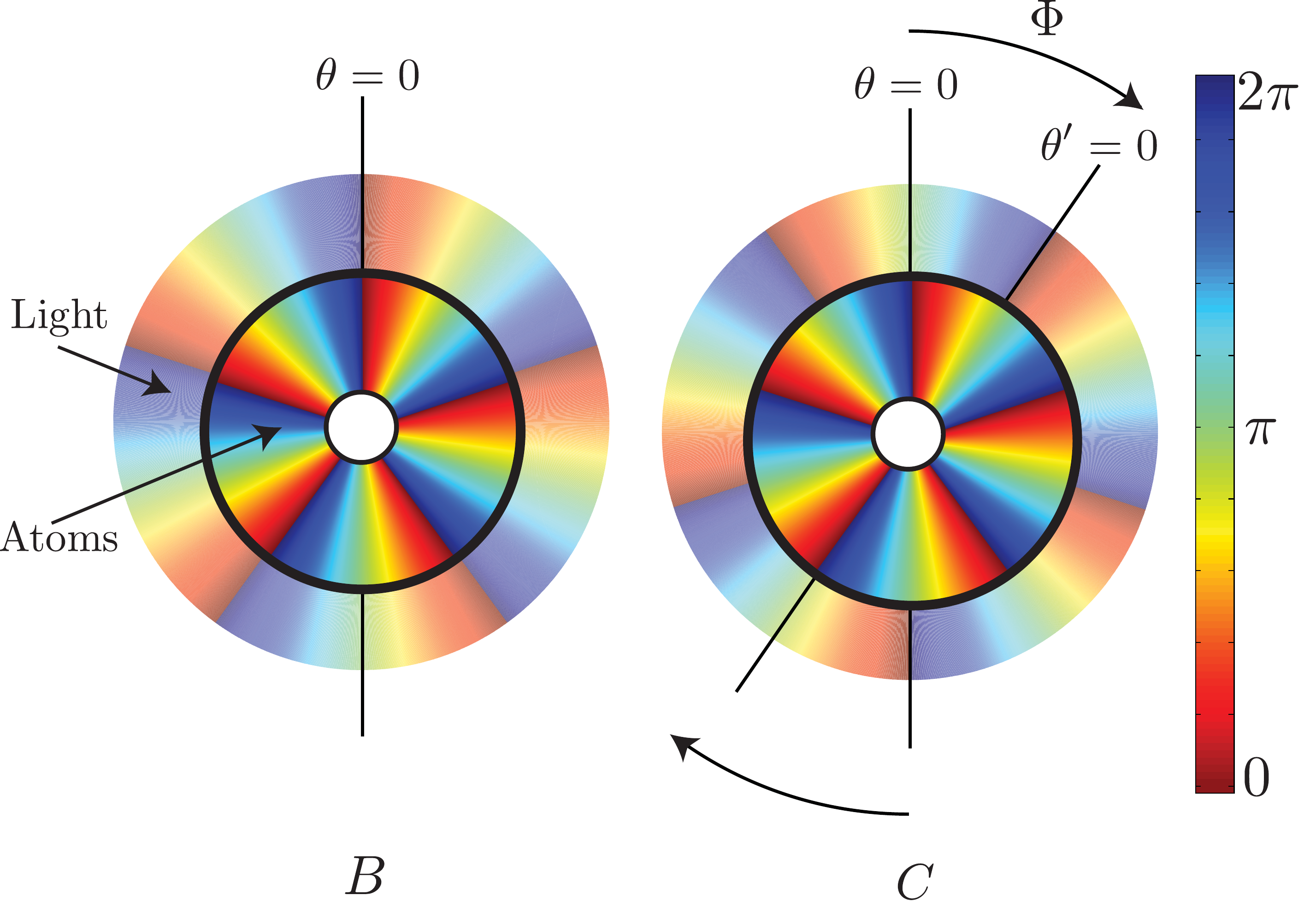}
\caption{(Color online). Schematic of rotation sensing using from counter-propagating atoms and Laguerre-Gauss beams. For illustrative purposes we show the $\ell=5$ case. Color represents relative phase of the LG beams (the outer ring) and the $m_{\rm F}=\pm 1$ atomic spin states (inner ring). The beams $B$ and $C$ correspond to the two $\pi/2$ pulses, shown in Fig.~\ref{fig:scheme}. The rotation of the LG beams causes a shift in the relative phase, while the atoms provide an inertial reference frame. As the atomic population difference after the second $\pi/2$ pulse (``C") depends on the relative phase of the LG beams \eqref{eq:Jzout}, the final population difference is sensitive to the rotation.  
}
\label{fig:LGraman}
\end{figure}

In our interferometer we couple  atoms in the $m_F = +1$ and $m_F = -1$ Zeeman levels, assumed to occupy motional states with centre of mass orbital angular momentum $+\hbar \ell$ and $-\hbar \ell$ respectively (Fig.~\ref{fig:energylevel}). Because this coupling conserves kinetic energy we set $\delta = 0$ as shown in Fig.~ \ref{fig:energylevel}(b). After preparation of the input states, a $\pi/2$ pulse is implemented by applying a Raman pulse of duration
\begin{equation} \label{eq:deltat}
t_{\pi/2}=\frac{\pi}{2} \frac{ \Delta}{\Lambda_0^2} \, ,
\end{equation}
such that
\begin{subequations}\label{eq:psi1}
\begin{align}
\hat{\psi}_{+1}(\textbf{r}, t_{\pi/2}) = & \frac{1}{\sqrt{2}}\left( \hat{\psi}_{+1}(\textbf{r}, 0) -i \hat{\psi}_{-1}(\textbf{r}, 0) e^{i 2 \ell \theta} \right), \\
\hat{\psi}_{-1}(\textbf{r}, t_{\pi/2}) = & \frac{1}{\sqrt{2}}\left( \hat{\psi}_{-1}(\textbf{r}, 0) -i \hat{\psi}_{+1}(\textbf{r}, 0) e^{-i 2 \ell \theta} \right),
\end{align}
\end{subequations}
where $\hat{\psi}_{j}(\textbf{r}, 0)$ is the Schr{\"o}dinger picture bosonic field operator for the $j$th spin state. The system then undergoes free evolution for some interrogation time $T$, during which an external rotation of the system will rotate the LG beams by an angle $\Phi = \int_{0}^T \Omega(t) dt$ relative to the inertial references provided by the counter-propagating BEC components, where $\Omega(t)$ is the angular frequency of the rotation. If the rotation is about the $z$ axis,  this is equivalent to shifting the coordinate system of the beams by some angle $\Phi$ to the rotated coordinate $\theta '=\theta+\Phi$.
This is also equivalent to a shift in the relative phase of the two LG beams of $\phi=2 \ell \Phi$. After the interrogation time the states are recombined with a second $\pi/2$ pulse, performed with the rotated LG beams. This is also described by Eq.~\eqref{eq:ramanHam} with LG beams given by Eq.~\eqref{eq:rabi} but now in terms of the rotated equatorial angular coordinate, $\Lambda_{\pm 1}(\theta ')$. At time $t_f= t_{\pi/2} + T +  t_{\pi/2}$ the field operators are 
\begin{subequations}\label{eq:psiout}
\begin{align}
 \hat{\psi}_{+1}(\textbf{r}, t_f) = & \frac{1}{2} \Big[ \left(1-e^{i \phi} \right) \hat{\psi}_{+1}(\textbf{r}, 0) \nonumber \\ 
 & -i e^{i 2 \ell \theta} \left(1+e^{i \phi} \right) \hat{\psi}_{-1}(\textbf{r}, 0)  \Big], \\ 
\hat{\psi}_{-1}(\textbf{r}, t_f) = & \frac{1}{2} \Big[ \left(1-e^{-i \phi}\right) \hat{\psi}_{-1}(\textbf{r}, 0) \nonumber \\ 
 & -i e^{-i 2 \ell \theta} \left(1+e^{-i \phi} \right) \hat{\psi}_{+1}(\textbf{r}, 0)  \Big] .
\end{align}
\end{subequations}
This final $\pi/2$ pulse acts to compare the relative phase of the Raman beams to the stationary phase-reference of the counter-propagating atomic modes, as  illustrated in Fig.~\ref{fig:LGraman}. In writing Eqs.~\eqref{eq:psiout} we have ignored the free-evolution of the atoms in the time between the two coupling pulses. We explore the effect of a finite period of free-evolution in Sec.~\ref{sec7}. Briefly, in the situation where only two motional eigenstates of the confining potential with equal and opposite angular momentum are occupied, then the relative phase due to the contribution from the kinetic energy cancels, and Eqs.~\eqref{eq:psiout} remains valid. 

This treatment assumes that the axis of rotation is perfectly aligned with  the axis of the ring trap. In the presence of a small off-axis contribution to the rotation, the accrued phase shift in Eqs.\eqref{eq:psiout} would be proportional to the $z$-component of the rotation only. A large off-axis contribution would cause a reduction in visibility due to the centre of the LG beam drifting relative to the centre of the ring-trap, reducing the overlap of the spatial profile of the atomic modes and the coupling profile defined by the LG beams. A slightly elliptical ring trap would have a similar effect. 


\subsection{Pseudo-Spin Description of Interference and Phase Sensitivity.} \label{sec2b}
The operator for the number difference  between the two Zeeman states is
\begin{equation}\label{eq:Jz}
\hat{J}_z(t)=\frac{1}{2} \left[\hat{N}_{+1}(t) - \hat{N}_{-1}(t) \right],
\end{equation}
where
\begin{equation}
\hat{N}_a(t) = \int d \textbf{r}\, \hat{\psi}_a^\dagger(\textbf{r}, t) \hat{\psi}_a(\textbf{r}, t),
\end{equation}
is the number operator for Zeeman level $m_F = a$.  By evaluating $\hat{J}_z$ at time $t_f$ [by substituting Eq.~\eqref{eq:psiout} into Eq.~\eqref{eq:Jz}] we see there are interference fringes  present in the number difference, and so this is the signal that can be used to measure the relative phase.  We find
%
\begin{equation}\label{eq:Jzout}
\hat{J}_z(t_f)=\hat{J}_x(0) \sin(\phi)-\hat{J}_z(0) \cos (\phi) \, ,
\end{equation}
where
\begin{equation}
\hat{J}_i=\frac{1}{2}\int d\textbf{r} \,\boldsymbol{\psi}^\dagger \sigma_i \boldsymbol{\psi} \, ,
\end{equation}
$\sigma_i$ is the $i$th Pauli matrix, and 
\begin{equation}
\boldsymbol{\psi}=
\begin{pmatrix} 
\hat{\psi}_{+1}(\textbf{r})  \\
\hat{\psi}_{-1}(\textbf{r}) e^{i 2 \ell \theta}
\end{pmatrix} \, .
\end{equation}
The $\{ \hat{J}_k \}$ operators obey the standard $SU(2)$ angular momentum commutation relations.
 We note that the $e^{i 2 \ell \theta}$ dependence in the definition of $\hat{J}_x$ and $\hat{J}_y$ comes from the $e^{i 2 \ell \theta}$ dependence in Eq.~\eqref{eq:psiout}, which is in turn a consequence of the use of LG beams in the Raman transitions. 

With $\hat{J}_z(t_f)$ as the signal, the corresponding phase uncertainty is
\begin{equation}\label{eq:deltaphi}
\Delta \phi = \frac{\sqrt{\mathrm{Var}\big[\hat{J}_z(t_f)\big]}}{\left|\partial_\phi \langle \hat{J}_z(t_f)\rangle \right|},
\end{equation}
which is smallest when $\phi=n \pi$ for integer $n$. For these values we find 
%
\begin{equation} \label{eq:deltaphinpi} 
\Delta \phi \bigg|_{\phi=n \pi}=\frac{\sqrt{\mathrm{Var}\left[\hat{J}_z(0) \right]}}{\left| \langle \hat{J}_x(0) \rangle \right|} = \frac{\xi}{\sqrt{N_t}} \, ,
\end{equation}
where $\hat{J}_k(0)$ are the pseudo-spin operators prior to evolution through the interferometer, $N_t$ is the total number of detected atoms,  and $\xi$ is the Wineland squeezing parameter \cite{Wineland1994},
\begin{equation} \label{eq:xidef} 
\xi = \frac{\sqrt{N_t \mathrm{Var}(\hat{J}_z)}}{J_{\perp}} \, .
\end{equation}
with the perpendicular spin length
\begin{equation} \label{eq:Jperp} 
J_{\perp} = \sqrt{\langle \hat{J}_x \rangle ^2+\langle \hat{J}_y \rangle ^2} \, .
\end{equation}
We note that for our choice of initial conditions  $\langle \hat{J}_y\rangle = 0$. Equation~(\ref{eq:xidef}) also takes into account the effect of atomic population in other angular momentum modes, which will have the effect of reducing the fringe contrast, which manifests itself as a reduction of $J_\perp$.

The definition of spin-squeezing is when  $\xi < 1$, which results in phase sensitivity beyond the SQL. In the next section we will discuss how this may be achieved with spin-exchange collisions.  The Wineland parameter essentially describes the metrological potential of a particular input state for a perfect rotation sensor, which is described by Eq.~\eqref{eq:Jzout}. It is unable to account for effects such as a finite interrogation time, or imperfections in a realistic interferometer, such as dephasing due to nonlinear interactions, or other dynamics within the interferometer which perturb the spatial profile of the wave-packets. The effects of these processes are analysed in Sec.~\ref{sec7}.

\section{Scheme for spin-squeezed rotation sensing} \label{sec2new}

\begin{figure*}
\includegraphics[width=0.8\textwidth]{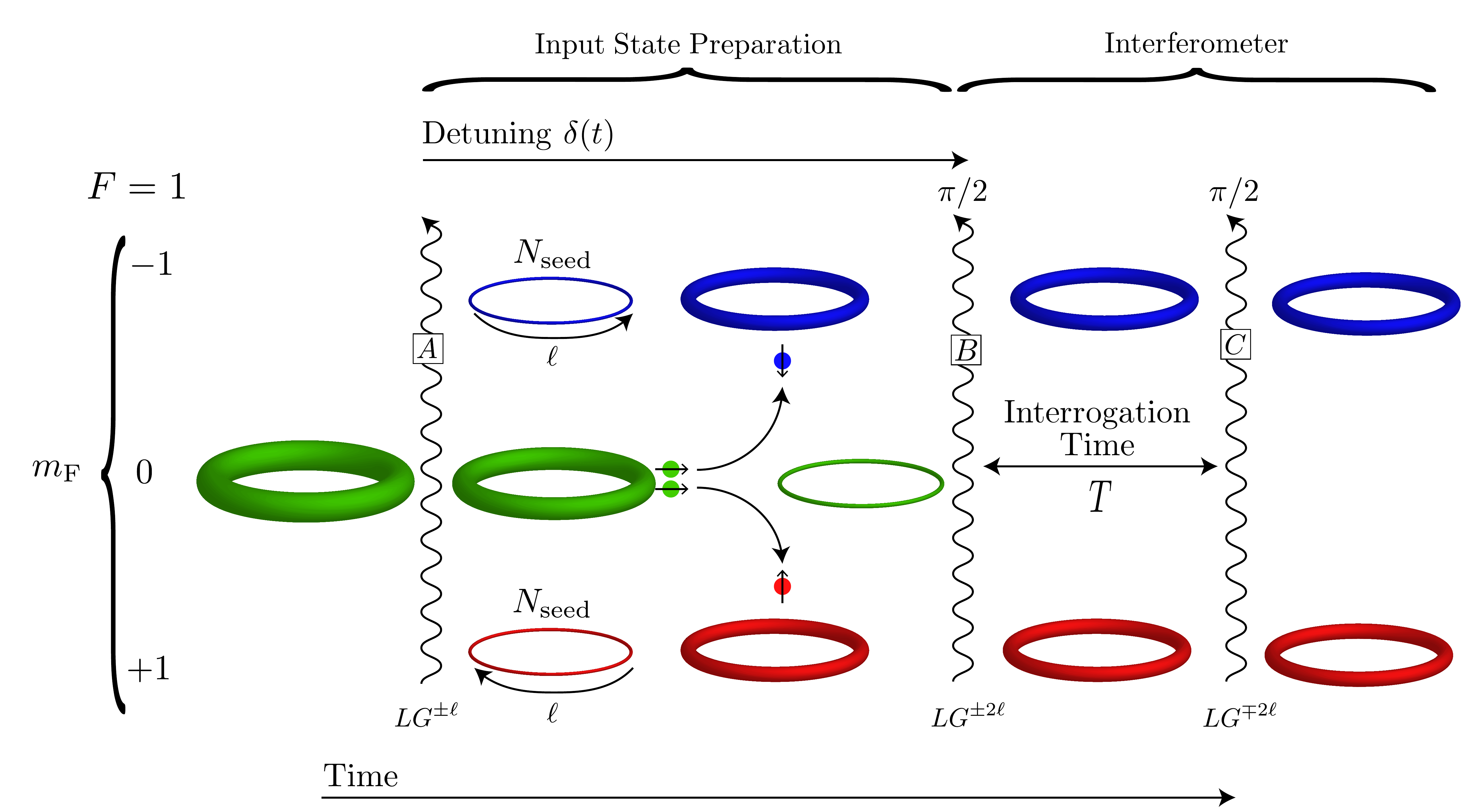}
\caption{(Color online). Schematic of the proposed spin-squeezed atom interferometer. A coherent seeding pulse (beam splitter A) transfers a small number of atoms $N_{\rm seed}$ from the 
$|0,0\rangle$
state to each of the $|\pm 1, \pm \ell\rangle$ states.
The quadratic Zeeman effect is used to cause resonant spin-exchange collisions by setting an appropriate bias magnetic field.
After the desired amount of population transfer, the system is tuned away from resonance and 
the trap is adiabatically relaxed, perhaps in the $z$ dimension, to reduce the effect of atomic interactions. The modes are mixed with a $\pi/2$ pulse (beam splitter B) which converts the relative number squeezing to phase squeezing. After accumulating a relative phase over interrogation time $T$ in the presence of a rotation, the 
$|\pm 1, \pm \ell\rangle$
modes are interfered with a final $\pi/2$ pulse (beam splitter C). Finally the number difference between the $m_{\rm F}=\pm1$ Zeeman states is measured, e.g by destructive imaging using a magnetic field gradient and Stern-Gerlach separation.}
\label{fig:scheme}
\end{figure*}

We now consider how to use quantum correlations generated from spin-changing collisions in a spin-1 BEC to enhance the sensitivity of the rotation sensor described in Sec.~\ref{sec2}. We build on the interferometry scheme presented in Sec. \ref{sec2} by using spin-exchange collisions between the Zeeman levels of this condensate to generate highly populated, monochromatic spin-squeezed input states, as illustrated in Fig.~\ref{fig:scheme}.  In summary:

\begin{enumerate}
\item A $^{87}$Rb spinor BEC is initially trapped in the $m_{\rm F}=0$ Zeeman level in an optical ring trap that can also  confine the $m_F=\pm 1$ states.

\item Two separate two-photon Raman transitions are used to coherently transfer a  ``seed" of atoms from $m_{\rm F}=0$ to $m_{\rm F}=\pm 1$, which will serve as the initial state for the subsequent spin changing dynamics \cite{Sengstock2004, Chang2005}.
The use of LG beams to implement the Raman transition also transfers orbital angular momentum to these spin states, such that the $m_{\rm F}=\pm1$ component acquires an orbital angular momentum of $\pm \ell \hbar$. 

\item The quadratic Zeeman effect is utilised to ensure spin-exchange collisions of atoms from the original condensate to the seeded modes are resonant.  These stimulated collisions rapidly increase the particle number in each mode without increasing the variance in the number difference, resulting in two highly monochromatic input states with a high degree of relative number squeezing, i.e the spin-squeezing parameter $\xi<1$ [Eq.~\eqref{eq:xidef}].

\item The trap is then adiabatically relaxed
to reduce the density and collision rate. A $\pi/2$ Raman pulse implemented by LG beams acts as a beam splitter to mix the two modes.  The system is then allowed to freely evolve for some interrogation time $T$ during which a rotation of the LG phase occurs relative to the phase reference provided by the rotating BEC components. A final $\pi/2$ pulse interferes the atoms and the number difference can be measured, which will depend on the rotation angle [Eq.~\eqref{eq:Jzout}]. The spin-squeezed input state allows this phase shift to be determined to beyond the precision allowed by the SQL.

\end{enumerate}

\subsection{Hamiltonian}\label{sec3}

To perform a rotation measurement with precision beyond the SQL, we wish to use relative number squeezed input states for use in the gyroscope described in Sec.~\ref{sec2}.  The initial state is an $F=1$ $^{87}$Rb spinor condensate in an optical ring trap with atoms in the $m_{\rm F}=0$ state. If the rotation $\Phi$ occurs only in the $z=0$ plane, and the radial profile of the optical LG mode is large compared to the trap radius $R$, then the operation of the interferometer is independent of the radial and axial degrees of freedom available to the atoms. Furthermore if the transverse confinement of the trap is sufficiently tight, it is reasonable to integrate out these dimensions. This affords us a 1D treatment of the system with position coordinate $\theta$, which will capture  the essential physics of the system. 

We write the atomic states as $|m_{\rm F},\ell\rangle$ where $m_{\rm F}$ is the electronic Zeeman sublevel and $\ell \hbar$ is the atomic centre of mass angular momentum mode occupied by the field (note that this is not the electronic orbital angular momentum quantum number). The full Hamiltonian describing the free evolution of a $F=1$ 1D spinor condensate (with Raman pulses) is \cite{Kawaguchi2012}
\begin{equation}\label{eq:spinHam}
\hat{\mathcal{H}}_{\mathrm{F}=1}= \hat{\mathcal{H}}_{T}+\hat{\mathcal{H}}_{\mathrm{SP}}+\hat{\mathcal{H}}_{\mathrm{SE}}+\hat{\mathcal{H}}_{\mathrm{Z}}+f(t) \hat{\mathcal{H}}_{\mathrm{R}},
\end{equation}
where $\hat{\mathcal{H}}_{\mathrm{R}}$ is Eq.~\eqref{eq:ramanHam} and $f(t)$ is some function of time which is either 1 or 0, whose purpose is simply to ``turn on" the Raman pulses at the appropriate times throughout the evolution.

The four other contributions to the Hamiltonian are the kinetic energy
\begin{equation}\label{eq:spinHamT}
\hat{\mathcal{H}}_{T}= \int d \theta \sum_{j=-1}^1 -\hat{\psi}^\dagger_j \frac{\hbar^2}{2mR^2}\frac{\partial ^2}{\partial \theta ^2} \hat{\psi}_j,
\end{equation}
the spin-preserving $s$-wave collisions
\begin{align}\label{eq:spinHamSP}
\hat{\mathcal{H}}_{\mathrm{SP}}=& \int d \theta \Big(\ \frac{c_0}{2}\hat{n}_0^2+ \frac{c_0+c_2}{2} \big[ \hat{n}_{+1}^2 + \hat{n}_{-1}^2 + 2 \hat{n}_0 \hat{n}_{+1} \nonumber \\
& +  2  \hat{n}_0 \hat{n}_{-1} \big] + (c_0-c_2) \hat{n}_{+1} \hat{n}_{-1} \Big),
\end{align}
the spin-exchange collisions
\begin{equation}\label{eq:spinexHam}
\hat{\mathcal{H}}_{\mathrm{SE}}= c_{2} \int d\theta \left( \hat{\psi}_{0}^{\dagger} \hat{\psi}_{0}^{\dagger} \hat{\psi}_{+1} \hat{\psi}_{-1} + \mathrm{H.c} \right),
\end{equation}
and the energy due to the quadratic Zeeman effect
\begin{equation}\label{eq:spinHamZ}
\hat{\mathcal{H}}_{\mathrm{Z}}=\hbar \delta_Z(t) \int d \theta \big( \hat{n}_{+1}+\hat{n}_{-1} \big).
\end{equation}
We have omitted the term describing the linear Zeeman effect as it can be eliminated by moving to the appropriate rotating frame.  In the above equations we have defined the number density operator $\hat{n}_k=\hat{\psi}^\dagger_k \hat{\psi}_k$, and the spin-independent and spin-dependent interaction constants $c_{0} = 2\hbar^2(2 a_{2} + a_{0})/(3RmA)$ and $c_{2} = 2\hbar^2(a_{2} - a_{0})/(3RmA)$ respectively. Note that $c_2<0$ for $^{87}$Rb. The transverse area due to integrating out two dimensions is $A$, $m$ is the mass of a $^{87}$Rb atom and $a_{S}$ is the scattering length for a collision process with final spin $S$. The term responsible for spin-squeezing  is Eq.~\eqref{eq:spinexHam}, which creates entangled atomic pairs by the Bose stimulated scattering of particles from the $m_{\rm F}=0$ states to  $m_{\rm F}=\pm 1$. We have also included an energy shift $\delta_Z(t)$ due to the quadratic Zeeman effect of each $m_{\rm F}=\pm 1$  levels relative to the $m_{\rm F}=0$ level.  This can be adjusted dynamically in an experiment by changing the strength of a bias magnetic field. 

\subsection{Seeding of input states}

\begin{figure}
\includegraphics[width=0.8\columnwidth]{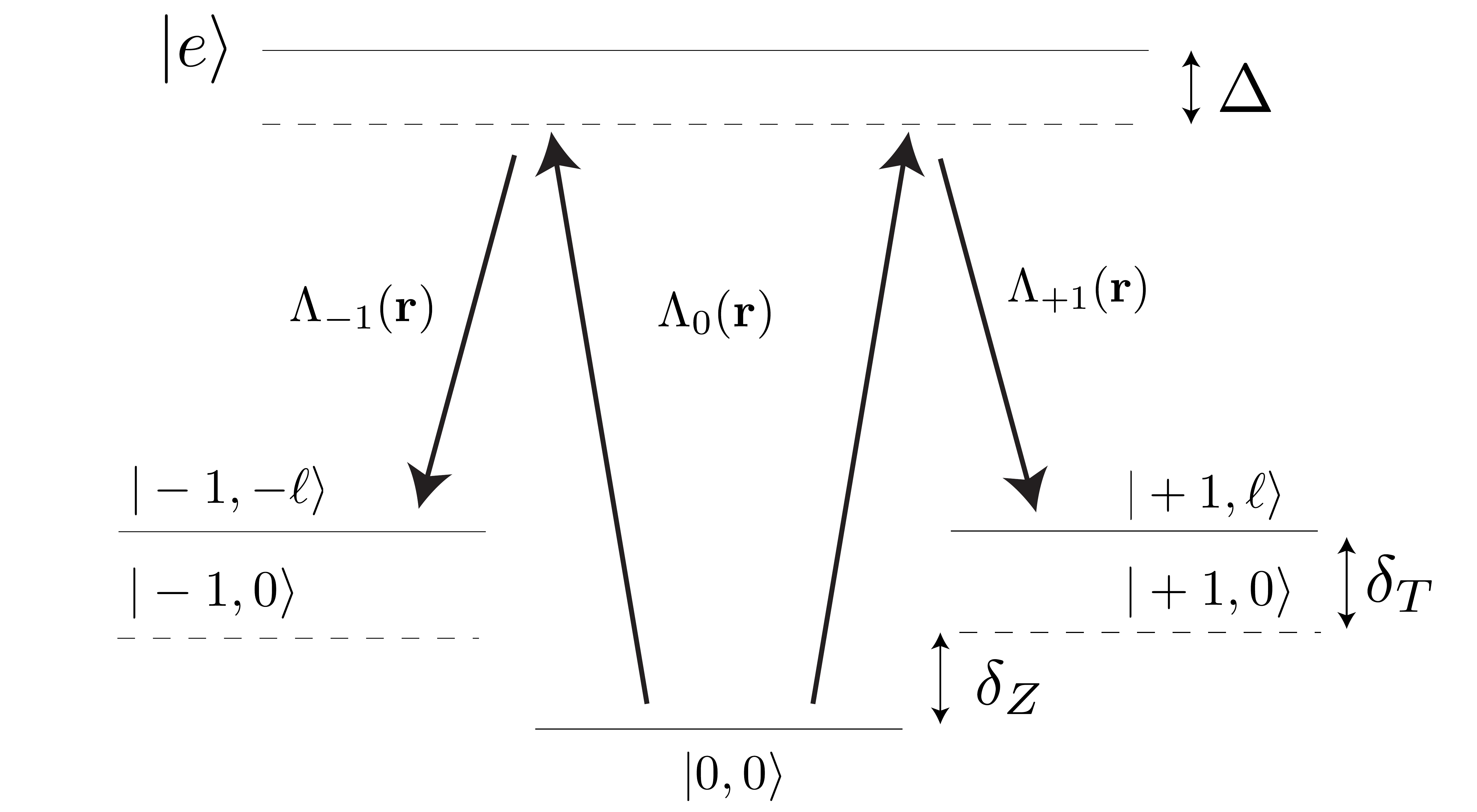}
\caption{Energy level diagram in presence of quadratic Zeeman $\delta_Z$ shift for two (simultaneous) seeding pulses, used to transfer a small number of atoms from the original $|0,0\rangle$ ground state to the $|\pm1,\pm \ell\rangle$ modes. This corresponds to pulses ``A" in Fig.~\ref{fig:scheme}. We include a two photon detuning $\delta_T=\hbar \ell^2/2mR^2$ to ensure the coupling process conserves kinetic energy. }
\label{fig:energylevelseed}
\end{figure}

To generate spin-squeezed input states, we first utilise two Raman transitions with LG beams with angular momentum $\pm \ell \hbar$ to coherently transfer a small fraction of the atoms, $|0,0\rangle \rightarrow | +1, + \ell \rangle$ and $|0,0\rangle \rightarrow | -1, - \ell \rangle$. The Hamiltonian for this process is given by  Eq.~\eqref{eq:ramanHam} with $a=0$ and $b=\pm 1$ with Rabi frequencies $\Lambda_0$ and $\Lambda_{\pm 1}(\theta)$ [Eq.~\eqref{eq:rabi}]. To ensure the transition is on resonance we choose the two photon detuning $\delta=\hbar \ell^2 /2mR^2$ i.e the kinetic energy given to the seed atoms by the Raman lasers. To create a seed of $N_{\rm seed}$ atoms in \emph{each} of the $|\pm 1, \pm \ell \rangle$ BEC components from an original $|0, 0\rangle$ condensate containing $N_0$ atoms, we use a pulse with duration
\begin{equation} \label{eq:seedpulse}
 t_{\mathrm{seed}}=\sqrt{\frac{N_{\rm seed}}{N_0}} \frac{\Delta}{\Lambda_0^2}.
\end{equation}

The seeding process creates the coherent initial state \cite{walls2008quantum},
\begin{subequations}  \label{eq:initialstate}
\begin{align}
|\psi \rangle&=|\alpha_0, \alpha_{+1}, \alpha_{-1} \rangle,  \\
&=  \mathcal{D}(\alpha_0) \mathcal{D}(\alpha_{+1}) \mathcal{D}(\alpha_{-1}) |0\rangle,
\end{align}
\end{subequations}
where
\begin{equation} \label{eq:displcedefn}
\mathcal{D}(\alpha_j)=e^{\alpha_{j} \hat{a}_{j}^\dagger-\alpha_{j}^* \hat{a}_{j}} ,
\end{equation}
with coherent amplitudes $\alpha_0=\sqrt{N_0}$, $\alpha_{+1}=-ie^{i \chi} \sqrt{N_{\rm seed}}$, $\alpha_{-1}=-ie^{i \chi} \sqrt{N_{\rm seed}}$. The single mode bosonic annihilation operators $\hat{a}_j$ are defined
\begin{subequations} \label{eq:coherentdefn}
\begin{align}
\hat{a}_0 &= \int d\theta \frac{\hat{\psi}_0}{\sqrt{2 \pi}},\\ 
\hat{a}_{\pm 1} &= \int d\theta \frac{\hat{\psi}_{\pm 1}}{\sqrt{2 \pi}}e^{\mp i\ell\theta}.
\end{align}
\end{subequations}
We allow for the possibility of a relative phase $\chi$ between the $|\pm{1},\pm \ell\rangle$ states and the original $|0,0\rangle$ coherent state, which could be imparted via a relative phase between the two LG beams.
To optimise the signal to noise ratio for our interferometer we choose $\chi$ such that maximum population growth is achieved.

\subsection{Spin-squeezing of input states}

Spontaneous spin-exchange collisions will naturally populate the $m_{\rm F}=\pm 1$ Zeeman states. The effect of the 
initial seeding allows for bosonically enhanced collisions to rapidly transfer correlated particles to the selected momentum modes of the interferometer.  We note however, that in a magnetic field regime where the quadratic Zeeman effect can be neglected that this collision process does not conserve kinetic energy --- the initial $|0,0\rangle$ state has no kinetic energy, whereas the seeded states have an energy $\hbar^2 \ell^2 /2mR^2$.  To allow the desired spin-exchange collisions to occur, we adjust the bias magnetic field and  utilise the quadratic Zeeman effect to make the collision
$|0,0\rangle + |0,0\rangle \rightarrow |+1,+\ell\rangle + |-1,-\ell\rangle$ resonant.  Of course, the undesired collision $|0,0\rangle + |0,0\rangle \rightarrow |-1,+\ell\rangle + |-1,+\ell\rangle$ is also resonant, but the seeding leading to bosonic enhancement will overwhelm this competing process.

 As the stimulated collisions populate the rotating modes, their mean-field energy increases [see Eq.~\eqref{eq:spinHamSP}] which also causes the spin-exchange collision process to move off-resonance. To keep the collision on resonance we adjust the bias magnetic field in the appropriate manner. Assuming the number density remains roughly uniform, the quadratic Zeeman energy required to ensure resonance at all times can be found by applying energy conservation
\begin{equation} \label{eq:deltaZ}
 \hbar \delta_Z(t)= E_0(t)-\frac{1}{2}[E_{+1}(t)+ E_{-1}(t) + 2 \hbar^2 \ell^2/2mR^2],
\end{equation}

where 
\begin{subequations} \label{eq:energies}
\begin{align}
E_0(t)/\hbar =& \frac{c_0}{L} N_0(t)+\frac{(c_0+c_2)}{L}N_{+1}(t) \nonumber\\
& +\frac{(c_0+c_2)}{L}N_{-1}(t),\\
E_{\pm 1}(t)/\hbar =& \frac{(c_0+c_2)}{L}N_0(t)+\frac{(c_0\pm c_2)}{L}N_{+1}(t) \nonumber\\
& +\frac{(c_0\mp c_2)}{L}N_{-1}(t),
\end{align}
\end{subequations}
are the mean-field energies, and $N_j= \int d\theta  \langle\hat{\psi}^\dagger_{j} \hat{\psi}_{j} \rangle$ is the expectation value of the number of atoms in the  $m_{\rm F}=j$   state. We find that this resonance condition is fairly robust, a relative error in $\delta_Z$ up to $10$ percent has negligible impact on the population transfer.

The result is two number-correlated counter-propagating matter-waves with equal but opposite angular momentum. With seeding, the protocol fails to create maximally number-correlated modes, i.e
\begin{equation} \label{eq:maxcorrelated}
\mathrm{Var}\left[ \hat{J}_z(r) \right] = N_{\rm seed}/4,
\end{equation}
rather than $\mathrm{Var}\left[ \hat{J}_z(r) \right] = 0$, as one would expect for any pairwise particle creation process. 
However, this can still be significantly less than $\mathrm{Var}\left[ \hat{J}_z \right] =(N_{+1}+N_{-1})/4$, which is the limit for uncorrelated modes. 

The bosonically-enhanced collisions into the desired angular momentum modes create highly monochromatic final states. After a sufficient number of atoms have been transferred to the desired modes via spin-exchange collisions, we perform the interferometry protocol described in Sec.~\ref{sec2}. To suppress further spin-changing collisions into the interferometer modes, we adjust the bias magnetic field such that $\delta_{Z} = 0$ and adiabatically relax the trap in the $z$ dimension to reduce the system density, while retaining an approximately 1D treatment of the system. This has the added advantage of minimising dephasing due to the spin preserving $s$-wave collisions, which are also reduced.

\section{Approximate analytic treatment of the maximum obtainable spin-squeezing}\label{sec4}

Analytic results that estimate the obtainable amount of  spin-squeezing can be found by making several simplifying approximations. In the following we assume all resonance conditions are met, and only consider the preparation of the input states.  

\subsection{Quantifying Spin-Squeezing} \label{sec4a}

The first approximation is assuming that the only significant population in the fields $\hat{\psi}_{\pm 1}$ is due to the seeded atoms in the $\pm \ell$  angular momentum modes.  In this situation we have
\begin{subequations}\label{eq:SM}
\begin{align}
\hat{\psi}_0 &\approx \frac{\hat{a}_0}{\sqrt{2\pi}},\\
\hat{\psi}_{\pm 1} &\approx  \frac{\hat{a}_{\pm 1}}{\sqrt{2\pi}}e^{\pm i \ell \theta},
\end{align}
\end{subequations}
with single mode bosonic annihilation operator $\hat{a}_k$, as in Eq.~\eqref{eq:coherentdefn}. 

The next approximation is that the initial $m_{\rm F}=0$ condensate is a large coherent state that remains essentially undepleted. The simplified Hamiltonian
\begin{equation}\label{eq:SMspinexHam}
\hat{\mathcal{H}}_{SE}=\frac{c_2 N_0}{L} \left(\hat{a}^\dagger_{+1} \hat{a}^\dagger_{-1} + \mathrm{H.c} \right),
\end{equation}
has the following solutions for the single mode bosonic operators $\hat{a}_{\pm 1}$:
\begin{subequations}\label{eq:analde}
\begin{align}
\hat{a}_{+1}(r) = & \hat{a}_{+1} \cosh (r) +i \hat{a}_{-1}^\dagger \sinh(r), \\
\hat{a}_{-1}(r) = & \hat{a}_{-1} \cosh (r) +i \hat{a}_{+1}^\dagger \sinh(r), 
\end{align}
\end{subequations}
where we have defined the squeezing parameter
\begin{equation} \label{eq:r}
r=- \frac{c_2N_0}{L} t_{\rm prep} \geq 0,
\end{equation}
and the time $t_{\rm prep}$ is the duration that the spin-exchange collisions are resonant. As the spin-exchange interaction strength $c_2$ is negative, $r$ is always positive. 

We take expectation values of the number operators for these modes with respect to the coherent states created by the seeding process Eq.~\eqref{eq:initialstate}. Only the phase $\chi$ of the $m_{\rm F}=\pm1$ states relative to the $m_{\rm F}=0$ state  has any physical consequence, so we are free to choose $\alpha_{\pm 1}$ to be real numbers with no loss of generality. The number of atoms created by the spin-exchange collisions in each Zeeman state is 
\begin{multline}\label{eq:Nanal}
N_{\pm 1} (r, \chi, N_{\rm seed}) = \sinh^2(r)+\big[\cosh(2r) \\
  -\sin(2\chi) \sinh(2r) \big] N_{\rm seed}.
\end{multline}
The unbounded exponential growth predicted here is an artifact of fixing $N_0$ in Eq.~\eqref{eq:SMspinexHam} and is often called the undepleted pump approximation.  It demonstrates the exponential increase in population due to bosonic enhancement created by seeding over the vacuum growth rate, which is  the term $N^{\rm vac}_{\pm 1}(r) = \sinh^2(r)$.  The result Eq.~\eqref{eq:Nanal} is only valid for $N_{\pm 1} (r, \chi) \ll N_0$.

The perpendicular spin length can be similarly evaluated
\begin{equation}\label{eq:spinlength}
J_{\perp}(r, \chi, N_{\rm seed})=\left| \cosh(2r)-\sin(2\chi) \sinh(2r) N_{\rm seed} \right|,
\end{equation}
which is simply $N_{\pm 1}(r, \chi)-N^{\mathrm{vac}}_{\pm 1}$, i.e the number of atoms transferred into the $m_{\rm F}=\pm 1$ states due to  the stimulated (rather than spontaneous) spin-exchange collisions.

\begin{figure}[h!]
\centering
\begin{subfigure}
\centering
\includegraphics[width=0.8\columnwidth]{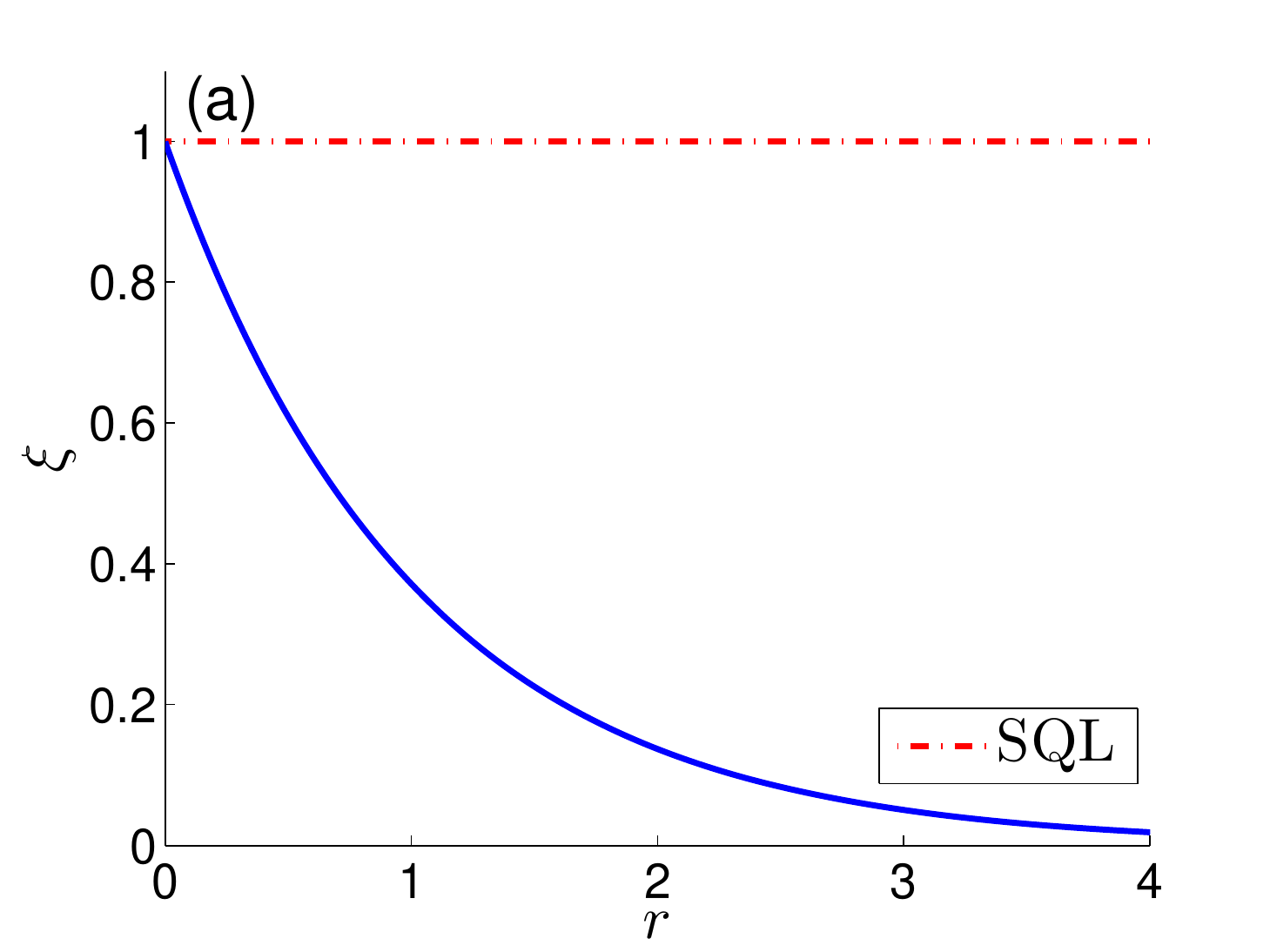}
\label{fig:4a}
\end{subfigure}
\hspace{2mm}
\begin{subfigure}
\centering
\includegraphics[width=0.8\columnwidth]{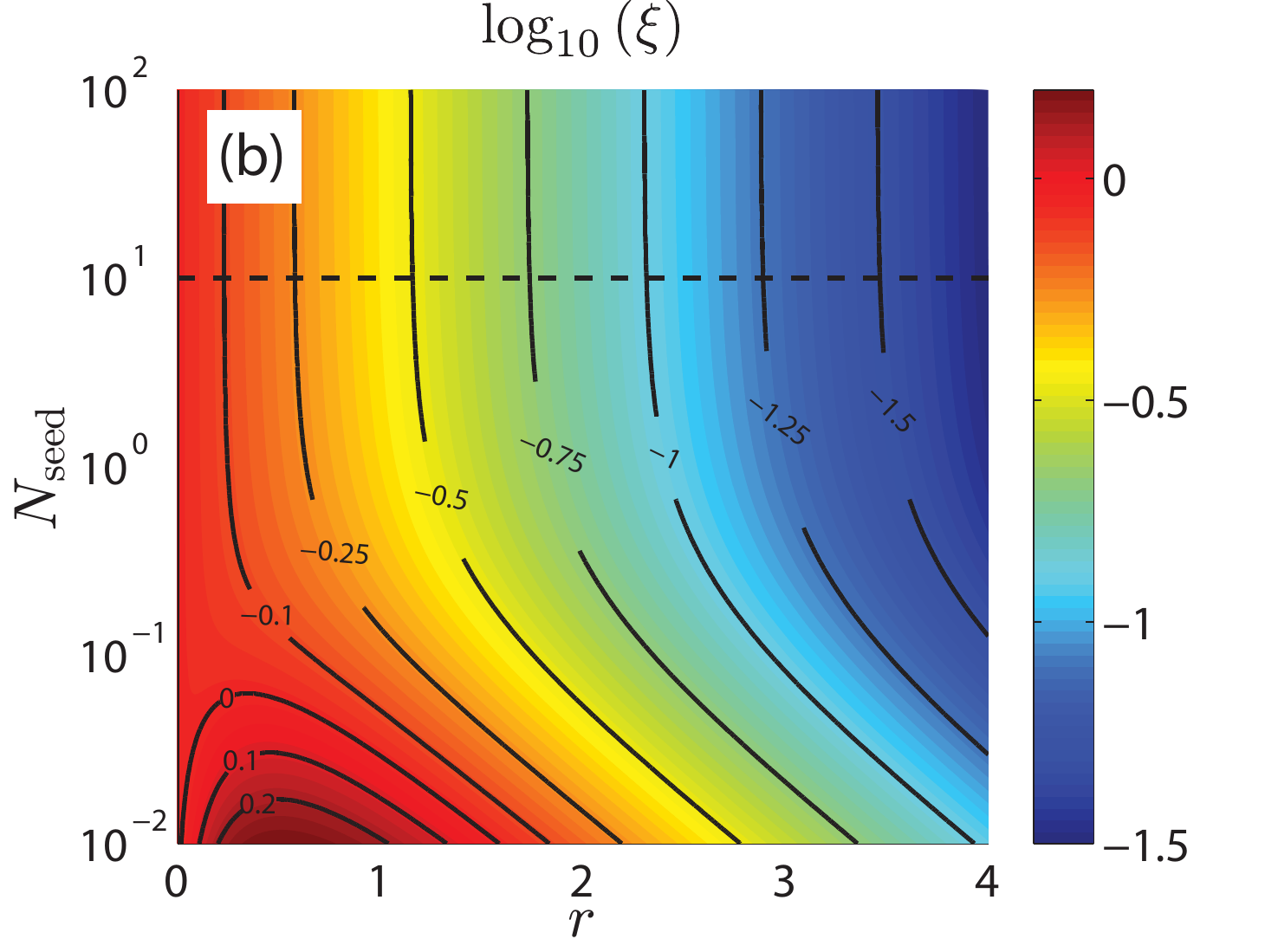}
\label{fig:4b}
\end{subfigure}
\qquad
\caption{(Color online). Analytic estimate of the spin-squeezing in the input states. (a) The spin-squeezing parameter $\xi(r, N_{\rm seed})$ is plotted as a function of squeezing parameter $r$ for $N_{\rm seed}=10$, which shows $\xi<1$ for $r>0$. (b) Contour plot showing $N_{\rm seed}$ dependence of $\xi$. We can see that $\xi$ is approximately independent of $ N_{\rm seed}$ for $N_{\rm seed} > 1$. In practice however, the $r$ at which the undepleted pump approximation breaks down depends on $N_{\rm seed}$, so there is effectively a maximum $r$ for each $N_{\rm seed}$. Dashed black line indicates contour plotted in (a).}
\label{fig:SQLxi}
\end{figure}

We are now in a position to evaluate the Wineland squeezing parameter Eq.~\eqref{eq:xidef} for spin-squeezed input states

\begin{align}\label{eq:squeeze}
\xi(r, \chi,N_{\rm seed})=& \sqrt{\frac{\frac{\sinh ^2(r)}{N_{\rm seed}}+\cosh (2 r)-\sin (2 \chi ) \sinh (2 r)}{\left[ \cosh (2 r)-\sin (2 \chi) \sinh (2 r) \right]^2}} ,
\end{align}
which is less than one for $r>0$. We find that $\xi$ and $N_{\pm 1}$ are minimised and maximised respectively for $\chi=3\pi/4$ radians, where we find
\begin{align}\label{eq:squeezeb}
\xi(r, \chi = 3\pi/4 ,N_{\rm seed}) =
\sqrt{\frac{e^{-4 r} \sinh ^2(r)}{N_{\rm seed}}+e^{-2 r}} ,
\end{align}
and henceforth we fix $\chi$ to this value. A plot of $\xi(r, N_{\rm seed})$ is shown in Fig.~\ref{fig:SQLxi} which demonstrates spin-squeezing. We can see that  for sufficiently small seed sizes vacuum growth  dominates, resulting in a short time where the system is spin anti-squeezed. It is straightforward to show that for $\xi<1$ we require
\begin{equation} \label{eq:xicond}
N_{\rm seed} > \frac{1}{2}e^{-3r} \sinh(r),
\end{equation}
for some amount of squeezing $r$. 

\subsection{Maximum  phase sensitivity} \label{sec4b}

\begin{figure}
\includegraphics[width=0.8\columnwidth]{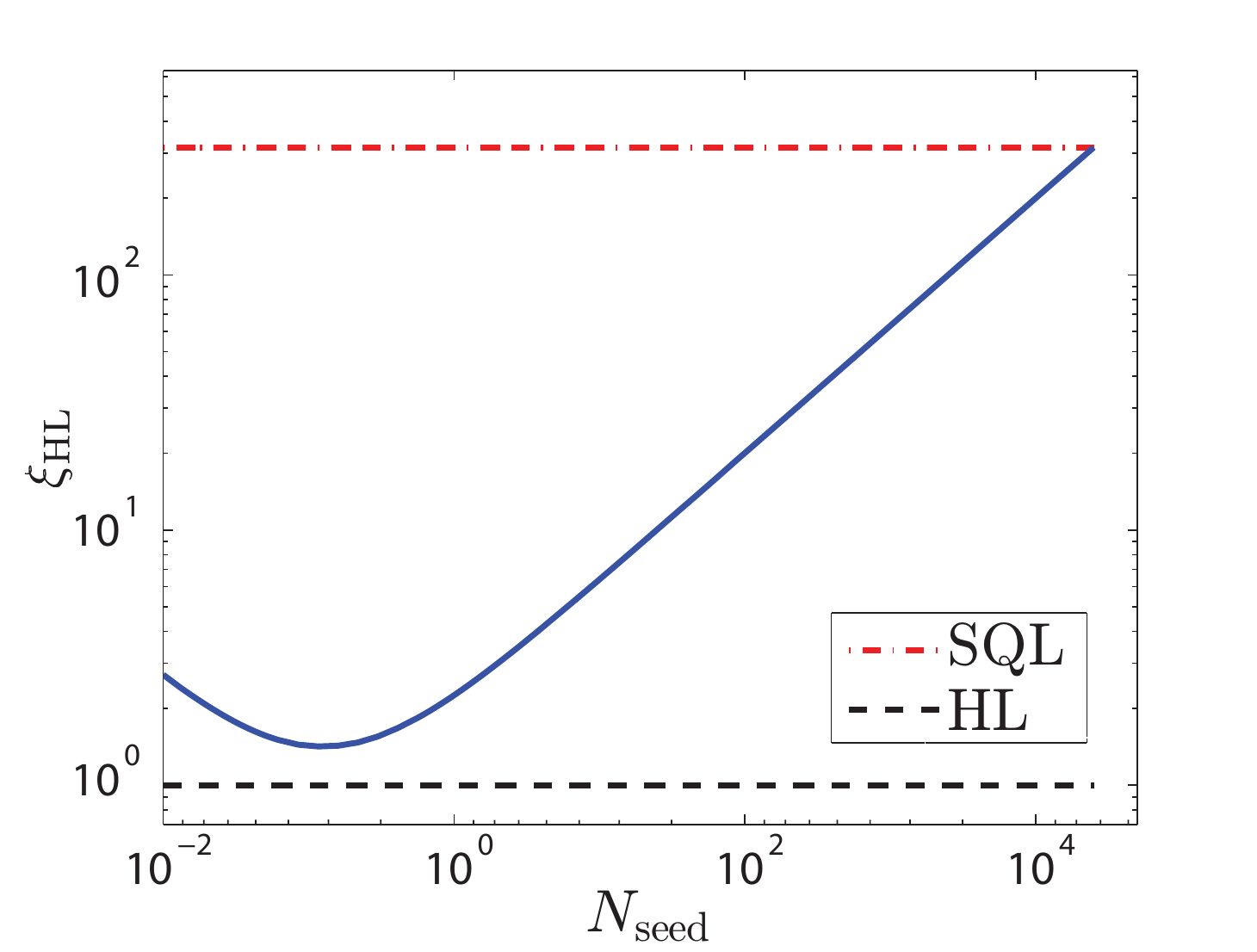}
\caption{(Color online). Spin-squeezing parameter normalised to the Heisenberg limit ($\xi_{\rm HL}$) for $N_t=10^5$ particles, i.e $N_0(t=0)=N_t$. Therefore the sensitivity for $N_{\rm seed}=10^5/2$ is  the SQL. The minimum occurs for a seed size of $N_{\rm seed}=1/4$ with a sensitivity $\sqrt{2}$ times the Heisenberg limit.}
\label{fig:xiplotty}
\end{figure}

We now calculate the maximum sensitivity for these input states. For a fixed number of total atoms $N_t$, the ultimate sensitivity attainable by any interferometer is the Heisenberg limit. This motivates us to examine $\xi_{\rm HL}=\sqrt{N_t} \xi$, which is Eq.~\eqref{eq:xidef} renormalised to the Heisenberg limit. We can evaluate $\xi_{\rm HL}$ analytically by setting $N_{\pm 1} (r, N_{\rm seed})=N_t/2$ [Eq.~\eqref{eq:Nanal}], and solving for the optimum $r$ we find
\begin{equation} \label{eq:optimumr}
r_{\rm opt} = \log \left(\sqrt{\frac{\sqrt{N_t (N_t+2)-4 N_{\rm seed}}+N_t+1}{4 N_{\rm seed}+1}}\right).
\end{equation}
Substituting this into $\xi_{\rm HL}$ eliminates $r$, but introduces dependence on $N_t$. The maximum sensitivity normalised to the Heisenberg limit is then
\begin{equation} \label{eq:HLxifull}
\xi_{\mathrm{HL}}=\frac{N_t(4 N_{\rm seed}+1)}{\sqrt{2N_{\rm seed}} \left(\sqrt{N_t (N_t+2)-4 N_{\rm seed}}+N_t+1\right)}.
\end{equation}
However for $N_t\gg2$, $\xi_{\rm HL}$ is approximately independent of $N_t$, and so we are able to obtain a simple expression for $\xi_{\rm HL}$ as a function of $N_{\rm seed}$ only. This is given by 
\begin{equation}\label{eq:HLxi}
\xi_{\mathrm{HL}} \approx \frac{1+4N_{\rm seed}}{\sqrt{8 N_{\rm seed}}},
\end{equation}
which is plotted in Fig.~\ref{fig:xiplotty}. 

Fig.~\ref{fig:xiplotty} demonstrates that the optimum sensitivity is achieved for small seeds. A decrease in sensitivity for seed sizes less than $ N_{\rm seed}=1/4$ is a result of using $\hat{J}_z$ as the signal. It is a well known result from quantum optics that the Heisenberg limit can be reached using squeezed vacuum ($N_{\rm seed}=0$) if $\hat{J}_z^2$ is analysed instead \cite{Kim1998, Lucke2011, Haine2015, Haine2015a}. Despite this, we chose to use a seed for the reasons of bosonic enhancement outlined in Sec.~\ref{sec3}. Additionally,  unseeded states are poorly suited to inertial measurement as the lack of coherent population makes the system insensitive to the inertial phase shift derived in Sec.~\ref{sec2}. In a more complete analysis the disadvantage of such small seed sizes is both the loss of signal contrast associated with diminished monochromacity, and the reduced  population in the squeezed spin states. It is therefore important to investigate the relationship between $\xi$ and seed size in the presence of depletion and full multi-mode dynamics. 

\section{Numerical determination of the Wineland  squeezing parameter for the  input states }\label{sec6}

The analytic results presented in Sec.~\ref{sec4} were derived using several  approximations.  Here we simulate the full dynamics of the fields by numerically solving for the dynamics using the truncated Wigner approximation (TWA). 

Briefly, the equation of motion for the Wigner function for the system can be found from the master equation by using correspondences between differential operators on the Wigner function and the original quantum operators~\cite{gardiner_zoller_book_04}.   By truncating all derivatives of third- and higher order (the truncated Wigner approximation), this is of the form of a Fokker-Planck equation that can then be sampled  by integrating trajectories of a Gross-Pitaevskii-like equation for the complex Wigner multi-mode phase space variables $\psi_k(\theta, \tau)$, with initial conditions stochastically sampled from the appropriate Wigner distribtion \cite{Johnsson2007, maviswigner}. The TWA equations of motion for the stochastic, complex fields used to reconstruct quantum expectation values are:
\begin{subequations}\label{eq:de}
\begin{align}
i& \frac{\partial}{\partial \tau}\psi_{\pm 1} =  \Bigg(-\frac{1}{2}\frac{\partial^2}{\partial \theta^2}  + \tilde{c}_{0} n +\tilde{c}_{2}\left(n_{\pm 1} + n_{0} - n_{\mp 1}\right)\Bigg)\psi_{\pm 1}\nonumber\\
 &+ \tilde{\delta}_Z(\tau) \psi_{\pm}+ \tilde{c}_{2} \psi_{\mp}^{*} \psi_{0}^2 + f(\tau) \Lambda_{\pm 1,0} \psi_{0} + f'(\tau) \Lambda_{\pm 1,\mp 1} \psi_{\mp 1} ,\\
i& \frac{\partial}{\partial \tau}\psi_{0} =  \left(-\frac{1}{2}\frac{\partial^2}{\partial \theta^2} + \tilde{c}_{0} n + \tilde{c}_{2} \left(n_{+1} + n_{-1}\right)\right)\psi_{0}\nonumber\\
 & + 2\tilde{c}_{2} \psi_{0}^{*} \psi_{+1}\psi_{-1} + f(t) \Lambda_{\pm 1,0}^{*} \psi_{\pm}. 
\end{align}
\end{subequations}
where $\Lambda_{i,j}=\Lambda^*_i \Lambda_j/\Delta$ are the coupling pulses between the components $i$ and $j$ with $\Lambda_{i,j} = \Lambda_{j,i}^{*}$, $n_j=|\psi_j(\theta, \tau)|^2$ and $n= n_{+1}+ n_{0}+ n_{-1}$. We introduce the dimensionless time coordinate $\tau=\omega t$ with $\omega=\hbar/m R^2$. Thus $\tilde{c}_S=c_S/\hbar \omega$ and $\tilde{\delta}_Z(\tau)=\delta_Z(\tau)/\omega$ are dimensionless. 

Expectation values of quantum observables are related to the complex Wigner variables by symmetric ordering
\begin{equation}
\left\langle \left\lbrace f\left( \hat{\psi}_k^\dagger(\theta, \tau), \hat{\psi}_k(\theta, \tau) \right) \right\rbrace \right\rangle=\overline{f\left( \psi_k^*(\theta, \tau), \psi_k(\theta, \tau) \right)} 
\end{equation}
where $\left\lbrace \right\rbrace$ denotes symmetric ordering \cite{walls2008quantum} and $\overline{f(\psi_k^*, \psi_k)}$ is an average over trajectories. 

For the purposes of spin-squeezing, the behaviour of the system is largely insensitive to the number statistics of the initial state \cite{Kitagawa1993}, so for simplicity we chose  a Glauber coherent state \cite{walls2008quantum}. It was shown in Ref.~\cite{Haine2009} that a mixture of coherent states with random phases, or equivalently, a Poissonian mixture of number states, behaves identically to a pure coherent state in this situation. Specifically, we chose the initial state of the system to be $\mathcal{D}(\alpha)|0\rangle$, with
\begin{equation}
\mathcal{D}(\alpha) = \exp\left(\alpha \hat{a}^\dag_g - \alpha^* \hat{a}_g\right) \, ,
\end{equation}
with 
\begin{equation}
\hat{a}_g = \int \Psi_0^*(\theta) \hat{\psi}_0(\theta) d \theta \, ,
\end{equation}
where $\Psi_0(\theta)$ is the unity normalised ground state of the system for all the atoms in the $m_{\rm F}=0$ component. As the potential is uniform, for all components this is $\Psi_j(\theta) = 1/\sqrt{2\pi}$. This corresponds to initial conditions for each TWA trajectory of
\begin{equation}
\psi_j(\theta,0) = \sqrt{N}_j\Psi_j(\theta) +\eta_j(\theta) \, ,
\end{equation}
where $N_{\pm1} = 0$ and $\eta_j(\theta)$ is Gaussian complex noise satisfying
\begin{equation}
\overline{\eta_i^*(\theta)\eta_j(\theta^\prime)} = \frac{1}{2}\delta_{ij}\delta(\theta-\theta^\prime) \, .
\end{equation}

For our TWA simulations we seed the $m_{\rm F}=\pm1$ Zeeman states in the $\ell=\pm 2$ angular momentum modes. We use a spatial grid with $M=16$ grid points and a sufficient number of stochastic trajectories to ensure that statistical error in the reconstructed expectation values is negligible. We consider a condensate initially with  $N_0(0)=10^5$ atoms in the $m_{\rm F}=0$ Zeeman level, which are confined to a trap of radius $R=15$  $\mu \mathrm{m}$ with transverse area $A=2.33 \ \mu \mathrm{m}^2$, which gives an effective 1D spin-dependent interaction strength of $\tilde{c}_2 = -6.82 \times 10^{-4}$. We use $s$-wave scattering lengths $a_0=110a_B$ and $a_2=107a_B$ \cite{Ho1998} where $a_B$ is the Bohr radius. 

In Sec.~\ref{sec4b} we found that under the approximations used to derive Eq.~\eqref{eq:SMspinexHam} our protocol has the largest quantum enhancement for a seed of $1/4$ atoms in each of the $m_{\rm F}=\pm 1$ modes with a sensitivity of $\xi_{\rm HL}=\sqrt{2}$. 
In a more realistic analysis with a multimode field this result is redundant as the unseeded modes contribute only to the noise in the signal. This situation favours a significantly larger seed, since bosonic enhancement of the atomic population in the seeded modes reduces sensitivity loss of this kind.

Additionally, in the absence of spin-squeezing the sensitivity of an interferometer increases with the number of atoms available for measurement, and while larger seed sizes reduce spin-squeezing, they do increase the amplitude of the interference fringes in $\hat{J}_z$ which boosts the signal to noise ratio. This means that while the system may be less squeezed for a larger seed, it could be significantly more sensitive overall.

For this reason we now investigate the dependence of the absolute sensitivity $\Delta \phi=\xi/\sqrt{N_t}$ on seed size, which depends on both the amount of spin-squeezing and the number of atoms $N_t$ used in the measurement. 
We compare the results of TWA simulations of Eq.~\eqref{eq:de} allowing for spontaneous scattering into multiple modes (MMTWA), and a single mode, three component model with necessarily perfect monochromacity, but otherwise retaining all features of the system (SMTWA). 

\begin{figure}[h!]
\centering
\begin{subfigure}
\centering
\includegraphics[width=\columnwidth]{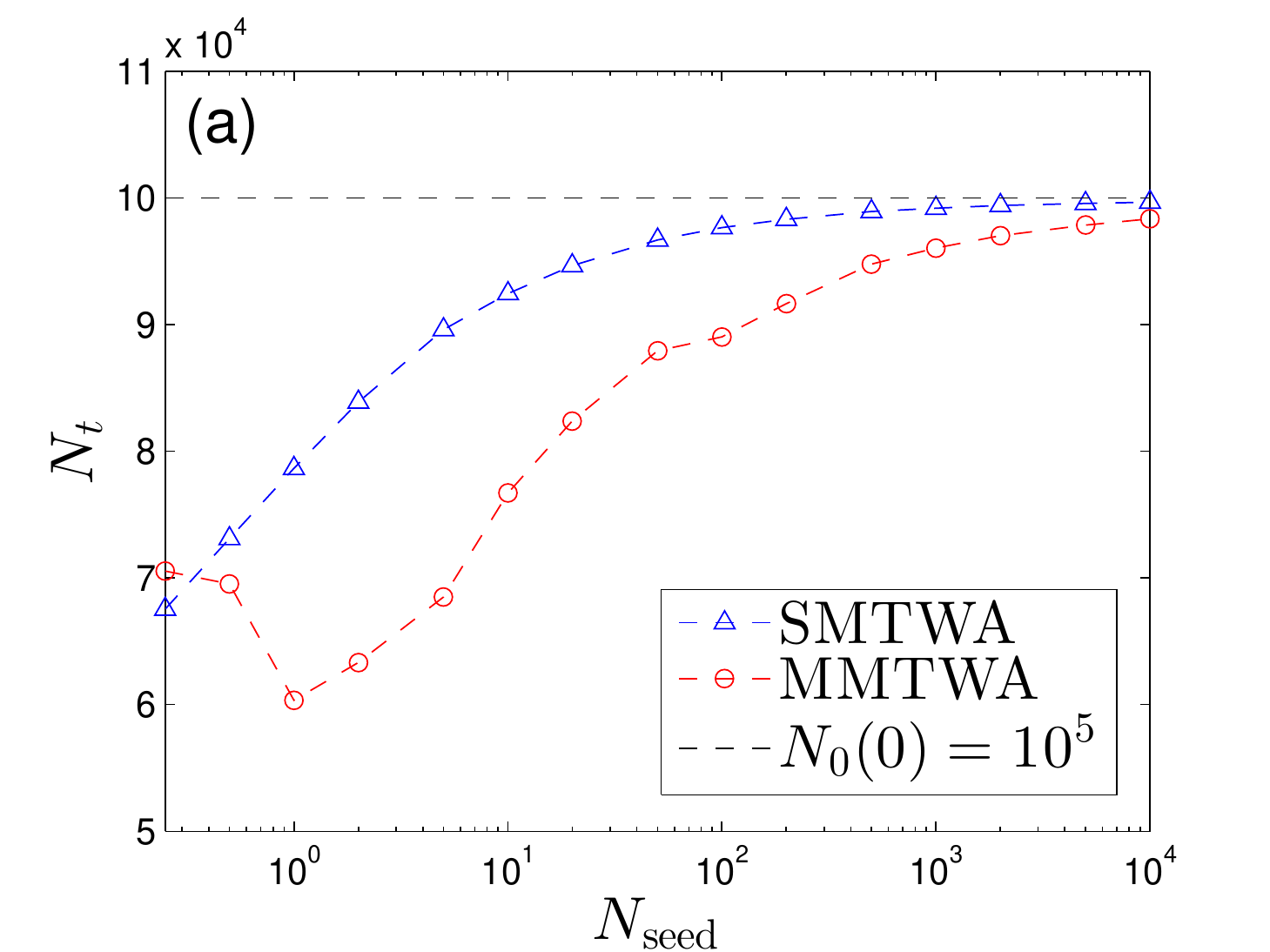}
\end{subfigure}
\hspace{2mm}
\begin{subfigure}
\centering
\includegraphics[width=\columnwidth]{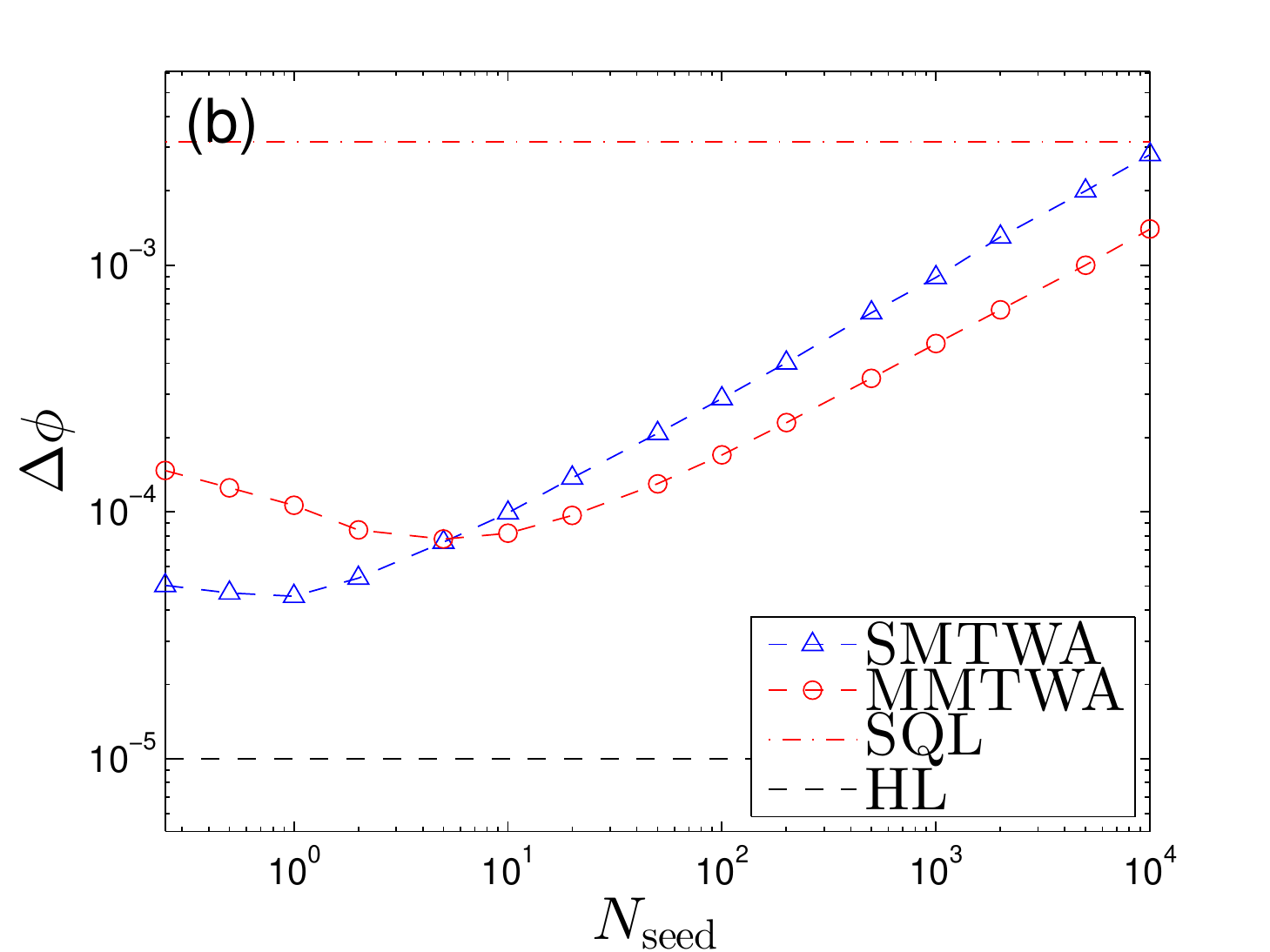}
\end{subfigure}
\qquad
\caption{(Color online). Comparison of the optimal characteristics of the single mode (SMTWA) and multimode (MMTWA) interferometer as a function of the seed size. (a) Optimum number of atoms transferred into the $m_{\rm F}=\pm 1$ Zeeman states via spin-exchange collisions for SMTWA, and the total number of $m_{\rm F}=\pm 1$ in the $\ell=\pm 2$ angular momentum modes for the MMTWA points. (b) The overall sensitivity $\Delta \phi=\xi/\sqrt{N_t}$, compared to the SQL and HL for a total number of $10^5$ atoms. The number in (a) is the value which optimises $\Delta \phi$. As each seed size has a different number of atoms available for measurement [shown in (a)] the SQL and HL is different for each point but for comparison we simply take the best case scenario ($N_t=10^5$) for each. Each point is the minimum sensitivity achieved for that seed size, i.e the points are for the optimum value of $\Delta \phi$, and the populations in (a) are the corresponding $N_t(r_\mathrm{opt})$.}
\label{fig:optr}
\end{figure}

These results are shown in Fig.~\ref{fig:optr}, and they demonstrate that  population in unseeded modes means that, for the largest absolute sensitivity, the optimum choice of seed size can be up to an order of magnitude larger than expected from the simplified analysis. The local minimum in the MMTWA curve in Fig.~\ref{fig:optr}(a) is caused by multimode effects, i.e the population of unseeded modes will grow independently of the seed size. For seeds that are too small, the population growth in the interferometer inputs is dominated by unseeded population growth, which results in a situation where a shorter preparation time (resulting in smaller $N_t$) is preferable.

\section{Simulation of the full interferometer sequence} \label{sec7}

The results presented so far have been concerned with the optimum preparation of an input state to the interferometer as described in Sec.~\ref{sec2new}. In deriving the Wineland parameter we have implicitly assumed that there is no evolution under Eq.~\ref{eq:spinHam} during the beam splitting with the Raman pulses, or during the interrogation time $T$. While the Raman pulses can be sufficiently fast that this is effectively true, a comparatively large interrogation time may be required to resolve a small rotation. During this period of free evolution, even in the absence of the nonlinearities in Eq.~\eqref{eq:spinHamSP} and Eq.~\eqref{eq:spinexHam}, there is a periodic oscillation in the higher order $\hat{J}_k$ moments which contribute to $\xi$. 

This behaviour can be understood qualitatively if we return to the undepleted pump approximation while retaining a multi-mode description of the field. This treatment is similar to the single mode analysis applied in Sec.~\ref{sec4a}, but including all  angular momentum modes 
\begin{subequations} \label{eq:psik}
\begin{align}
\hat{\psi}_{+1}=& \sum^\infty_{k=- \infty} \frac{\hat{a}_k}{\sqrt{2 \pi}} e^{i k \theta} , \\
\hat{\psi}_{-1}=& \sum^\infty_{j=- \infty} \frac{\hat{b}_j}{\sqrt{2 \pi}} e^{i j \theta} .
\end{align}
\end{subequations}
We have introduced $\hat{a}_{+1}=\hat{a}$ and $\hat{a}_{-1}=\hat{b}$ to avoid confusion with subscripts $j,k$, which are the integers that label the angular momentum modes of the trap. 

The signal at the output of the interferometer is $\hat{J}_z(t_f)$.  To relate this to the quantity of interest during the interrogation time we transform it backwards in time through the final beam-splitter and find
\begin{equation} \label{eq:backevolve}
e^{-i \hat{J}_x \pi/2} \hat{J}_z(t_f) e^{i \hat{J}_x \pi/2} = \hat{J}_y(T).
\end{equation}
This indicates that the quantum statistics of the signal at the output are related to $\hat{J}_y(T)$. Therefore we require $\hat{J}_y(T)$ to be squeezed after some interrogation time $T$ to achieve $\xi<1$. After the first $\pi/2$ pulse, i.e at the beginning of the interrogation time, the initial state is
\begin{equation} \label{eq:stateT}
|\psi(T=0) \rangle = e^{-i \hat{J}_x \pi/2}  e^{-i \hat{\mathcal{H}}_{\mathrm{SE}} \tau_{\mathrm{prep}}} e^{-i \hat{\mathcal{H}}_{\mathrm{Seed}} \tau_{\mathrm{seed}}}| 0 \rangle
\end{equation}
where $\mathcal{H}_{\mathrm{Seed}}$ is the seeding pulse, i.e Eq.~\eqref{eq:ramanHam} with  $i=0$ and $j=\pm 1$. In the basis of Eq.~\eqref{eq:psik} the time dependence in $J_k$ due to kinetic energy during the interrogation time is easily evaluated by substituting the operators
\begin{subequations} \label{eq:psikt}
\begin{align}
\hat{a}_k(T)=& \hat{a}_k(T=0)e^{-i k^2 \omega T/2}, \\
\hat{b}_j(T)=& \hat{b}_k(T=0)e^{-i j^2 \omega T/2}, 
\end{align}
\end{subequations}
into $\hat{J}_k$ (expanded into this basis), and by exploiting the orthogonality of the angular momentum modes
\begin{equation} \label{eq:deltafn}
\delta_{j,k}=\frac{1}{2 \pi} \int_0^{2 \pi} e^{-i k \theta} e^{i j \theta} d\theta.
\end{equation}
This gives operators of the form
\begin{multline} \label{eq:JyT}
\hat{J}_y(T)= \sum^\infty_{k=- \infty}  \hat{A}_k \cos[(2k\ell-2\ell^2)\omega T] \\
 + \hat{B}_k  \sin[(2k\ell-2\ell^2)\omega T],
\end{multline}
where we have defined the summands
\begin{subequations} \label{eq:AandB}
\begin{align}
\hat{A}_k =& \frac{i}{2} \Big( \hat{a}_k^\dagger \hat{b}_{(k-2\ell)} - \mathrm{H.c} \Big), \\
\hat{B}_k =& \frac{1}{2} \Big( \hat{a}_k^\dagger \hat{b}_{(k-2\ell)} + \mathrm{H.c} \Big),
\end{align}
\end{subequations}
which satisfy $\hat{J}_y(0)=\sum_k \hat{A}_k$ and $\hat{J}_x(0)=\sum_k \hat{B}_k$. The expectation values of these terms in the summand with respect to the state Eq.~\eqref{eq:stateT} are $\langle \hat{A}_{k} \rangle = 0$, $\langle \hat{B}_{k} \rangle = N_k^{\mathrm{coh}}$ where $N_k^{\mathrm{coh}}$ is the coherent population in the $k$th angular momentum mode, which is 0 for the unseeded modes ($k \neq \ell$). Clearly, when $k=\ell$ the time dependence vanishes from Eq.~\eqref{eq:JyT}. Thus it is apparent that only unseeded modes could possibly contribute to any dynamics, and so $\langle \hat{J}_y \rangle$ is static.  A similar argument gives the same conclusion for $\langle \hat{J}_x \rangle$ and $\langle \hat{J}_z \rangle$.

\begin{figure}
\includegraphics[width=\columnwidth]{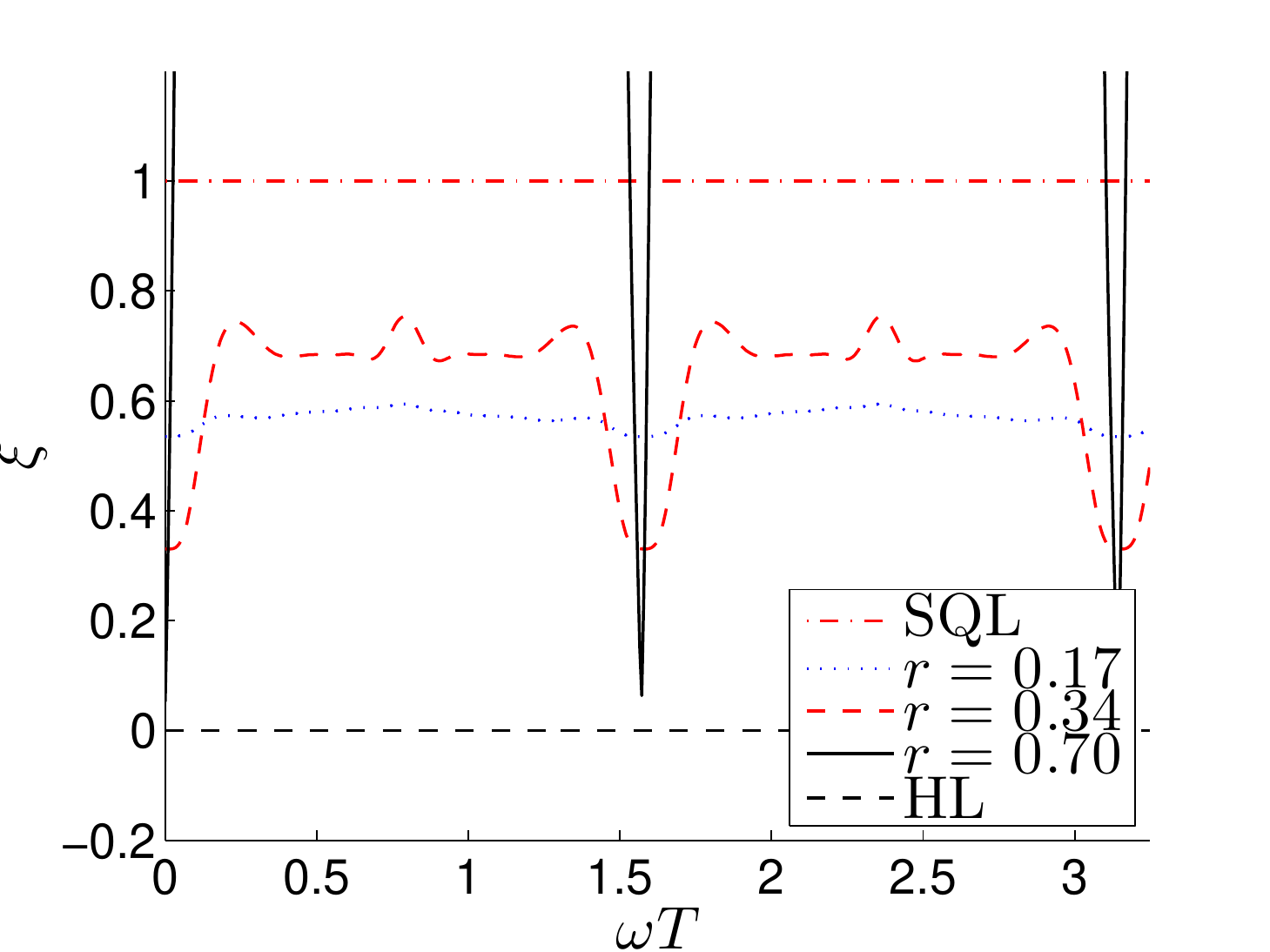}
\caption{(Color online). Dynamics of the Wineland parameter during the interrogation time. The solid black curve is close to maximally squeezed, with a large preparation time and a small initial seed. The dashed blue curve is only weakly squeezed, with a larger seed and shorter preparation time and demonstrates only mild dynamics. The dot-dashed red indicates an intermediate regime between the two extremes. The parameters used were $\tau_{\mathrm{prep}}/\omega = \{125, 60, 30\}$~ms, with seed sizes of $N_{\rm seed} = \{100, 5000, 10 000\}$ respectively. The squeezing parameter is calculated from Eq.~\eqref{eq:r}. Nonlinear interactions during the interrogation time further complicate this analysis, and so for illustrative purposes we have considered the case where the transverse area $A$ has been increased such that the nonlinear interactions are negligible, i.e $\tilde{c}_2, \tilde{c}_0=0$.  We consider the situation whereby the transverse confinement is such that the interaction cannot be ignored in Sec.~\ref{sec8}.}
\label{fig:xidynamic}
\end{figure}

During the interrogation time $\xi$ depends on the variance of $\hat{J}_y$. The expectation value of $\langle \hat{J}_y^2 \rangle$ contains a large number of terms. While the symmetry of the initial state ensures that many of these vanish, several survive and contribute to the time dependence of $\langle \hat{J}_y^2 \rangle$. One such term is
\begin{equation} \label{eq:dynamic}
\langle \hat{A}_k \hat{A}_{k} \rangle = \frac{1}{8} \sinh^2 \left(2 r_k \right),
\end{equation}
which carries a factor $\propto \cos^2[(2k\ell-2\ell^2)\omega T]$. To illustrate the effects of such dynamic terms on the system, Fig.~\ref{fig:xidynamic} shows the Wineland parameter as a function of interrogation time $T$. As expected from Eq.~\eqref{eq:dynamic}, the harder the system is squeezed the more sharply the Wineland parameter dips, and the lower the dip. Importantly, there are periodic revivals which indicate times when the system is optimally squeezed. However, these revivals do not necessarily indicate the optimum time for measurement, which we explore in Sec.~\ref{sec8}.

\section{Optimum Rotation Sensitivity} \label{sec8}

In this section we consider the overall performance of the interferometer in measuring rotations given a fixed initial atom number in the BEC.  The optimum time to perform a rotation measurement in a spin-squeezed system is not necessarily when the Wineland parameter $\xi$ is minimised. While increasing the interrogation time increases the sensitivity of a rotation measurement, dephasing due to atomic interactions can rapidly destroy the signal to noise ratio. To investigate the relationship between rotation sensitivity $\Delta \Omega$ and interrogation time we assume a constant rotation of the form $\Phi = \Omega T$. As outlined in Sec.~\ref{sec2}, the relative phase accumulated between the counter-propagating $m_{\rm F} = \pm 1$ components as a result of this rotation is $\phi = 2 \ell \Omega T$. The rotation uncertainty is related to the phase uncertainty by
\begin{equation} \label{eq:deltaomega}
\Delta \Omega(T) = \frac{\Delta \phi (T)}{2 \ell T},
\end{equation}
where $\Delta \phi$ is given by Eq.~\eqref{eq:deltaphi}. 

\begin{figure}
\includegraphics[width=0.8\columnwidth]{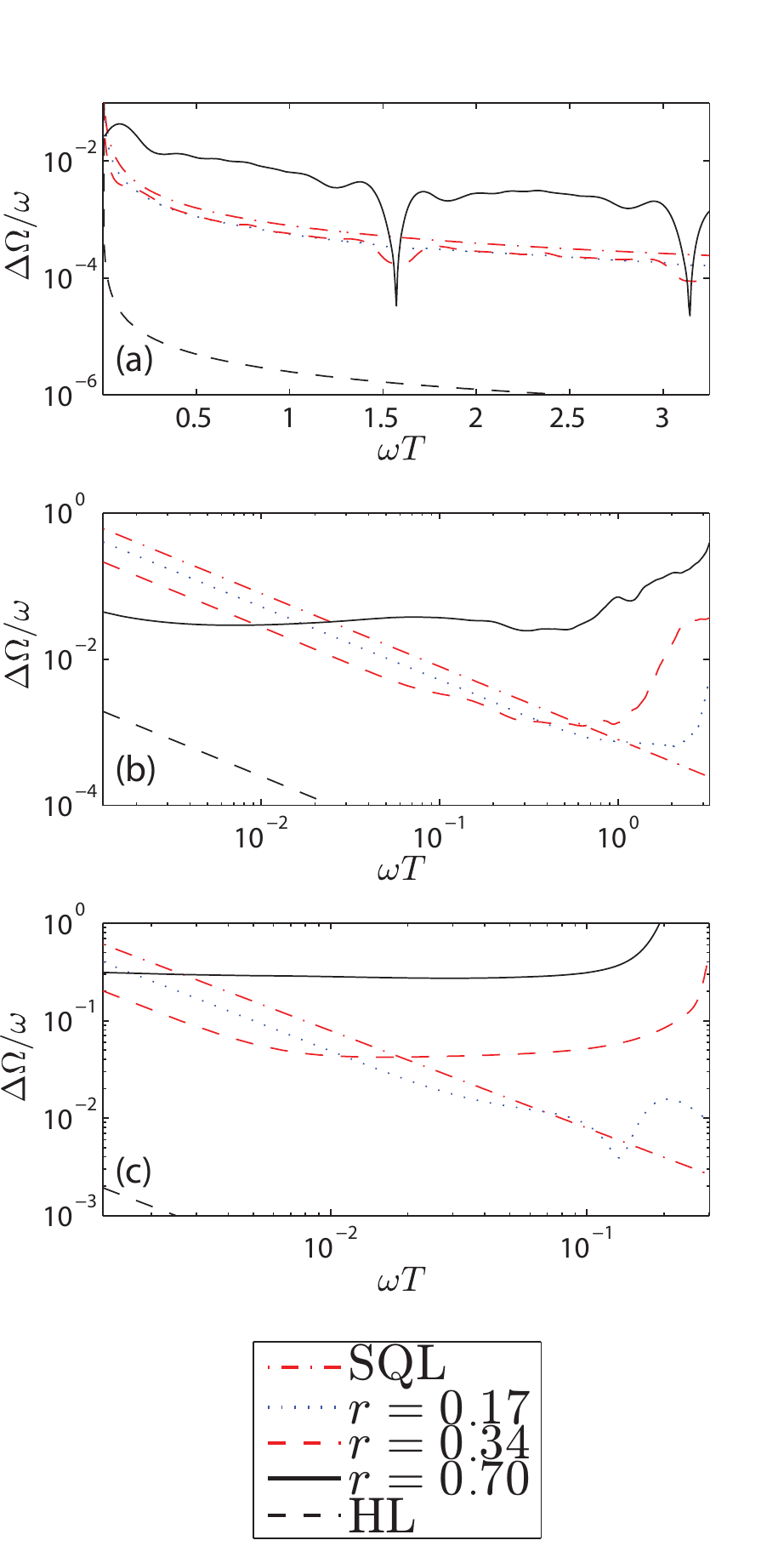}
\caption{(Color online). Absolute uncertainty $\Delta \Omega$ in the rotation measurement as a function of the interrogation time $T$ for effective 1D spin-dependent interaction strengths of (a) $\delta \tilde{c}_2=0$, (b) $ \delta \tilde{c}_2=0.02$ and (c) $\delta \tilde{c}_2=1$ during the interrogation time. We have defined the ratio $\delta \tilde{c}_2=\tilde{c}_2(T)/\tilde{c}_2(0)$ of the interaction strength during the interrogation time $\tilde{c}_2(T)$ to the interaction strength during the preparation time $\tilde{c}_2(0)$. These depend on the transverse area $A$, which can be controlled by relaxing the confinement in the $z$ direction. Even for small trapping frequencies the rotation sensitivity rapidly becomes worse than the SQL, and so (b) and (c) are plotted on a logarithmic time scale.}
\label{fig:deltaomega}
\end{figure}

Due to vacuum growth [the $\sinh^2(r)$ term in Eq.~\eqref{eq:Nanal}], smaller seed sizes with longer preparation times will have a higher fraction of atoms in the unseeded angular momentum modes. In turn, this means that atomic interactions during the interrogation time play a more significant role in determining a suitable regime for optimum $\Delta \Omega$. For this reason we would no longer expect a small seed which is highly squeezed to be optimal, as was indicated in Fig.~\ref{fig:optr}. To demonstrate this, Fig.~\ref{fig:deltaomega} shows $\Delta \Omega$ as a function of $T$ for a variety of atomic interaction strengths (parameterised by the transverse area $A$) and degrees of squeezing.

As expected Fig.~\ref{fig:deltaomega}(a) indicates that with no interactions the revivals present in Fig.~\ref{fig:xidynamic} are still present and represent optimum measurement times, and the sensitivity improves further if later revivals are used. However, for non-zero atomic interactions the optimum measurement time is largely independent of these revivals, and instead depends almost exclusively on the relative strength of the interactions.  This is related to both the population in the unseeded modes relative to the coherent seeded population, as well as the density. 
Figs.~\ref{fig:deltaomega}(b -- c) show the optimum seed size and preparation times  depend strongly on the interrogation time. In an attempt to minimise the effect of this de-phasing, we added a $\pi$ pulse at $t=T/2$. However, we found that this did nothing to recover the initial rotation sensitivity.

The deleterious effect of interactions shown in Fig.~\ref{fig:deltaomega} can be minimised by increasing the transverse area $A$. In order to maintain a 1D geometry and uniform radial density the transverse confinement should remain tight, but the trap could be relaxed in the $z$ dimension.

\section{Conclusions} \label{sec9}

We have analysed in detail the performance of a spin-squeezed rotation sensor based on a spin-1 Bose-Einstein condensate.  The spin-squeezed input states are generated via Bose-stimulated spin exchange collisions, following a coherent seed. Despite the fact that the spin-exchange Hamiltonian can reach the HL for a vacuum initial state, a seed must be used for rotation sensing as otherwise there is no coherent population to break the symmetry in the initial state to distinguish between clockwise and counter-clockwise rotations. The uniform density of the BEC components in the ring geometry gives good overlap at all times, and dynamically adjusting the quadratic Zeeman effect allows the generation of  a large, highly number-squeezed  input state for the interferometer that is potentially well-suited for rotation sensing. 

Considering only the preparation of the input state, we found that a small seed (of the order of $N_{\rm seed} /N_0 \approx 10^{-4}$) is able to achieve optimum spin-squeezing, as shown in Fig.~\ref{fig:optr}(b).  However, during the interrogation time, the beating between the incoherent population in the unseeded modes makes a highly squeezed regime optimal only if measurements are performed at specific times. Measurement sensitivity could be  enhanced by increasing $\ell$, which we have not considered in this work. The rotation sensitivity increases with $\ell$, which is  apparent from inspecting Eq.~\eqref{eq:deltaomega}, although there will be technical limitations on the maximum orbital angular momentum of the Laguerre-Gauss optical beams.

When the system is sufficiently dilute the effects of collisions can be small.  In this situation the optimal measurement times are in a narrow window where the Wineland squeezing parameter revives. However, for any significant collisional interactions the optimal measurement time is relatively short such that the effect of phase-diffusion is small. Such small interrogation times may be undesirable for a precision measurement of a rotation, as indicated by Eq.~\eqref{eq:deltaomega}. Performing the interferometry in a sufficiently dilute regime would reduce the deleterious effects of interactions, and allow for significant reduction in shot noise, and increased interrogation times, as is usually done in precision atomic interfometry gyroscope experiments \cite{Gustavson1997, Gustavson2000, Durfee2006}.  The rapid dephasing and highly dynamic sensitivity indicate that it may be favourable to consider methods other than atomic interactions to generate spin-squeezing. Another possible avenue for quantum-enhanced sensing is atomic-photon hybrid techniques \cite{Hammerer2010, Sewell2014, Szigeti2014, Chen2015, Szigeti2014}, which may avoid some of the complicating factors of this work.

\begin{acknowledgments}

Many of the numerical simulations were performed using XMDS2 \cite{Dennis2013} on the University of Queensland School of Mathematics and Physics high performance computing cluster `Obelix'.  We thank Leslie Elliot and Ian Mortimer for computing support. The authors would also like to thank Stuart Szigeti and Daniel Linnemann for useful discussion and feedback. SAH acknowledges the support of Australian Research Council Discovery Early Career Research Award DE130100575. JS was supported through the Australian Research Council Centre of Excellence for Engineered Quantum Systems (Centre CE1101013), whilst MWJB was supported by an Australian Research Council Future Fellowship (Project FT100100905). MJD acknowledges the support of Australian Research Council Discovery Project DP1094025 and the JILA Visiting Fellows program.
\end{acknowledgments}

\bibliographystyle{apsrev4-1}
\bibliography{mybib_final}

\end{document}